\documentclass[sigconf,nonacm]{acmart}

\AtBeginDocument{%
  }

\setcopyright{acmlicensed}
\copyrightyear{2018}
\acmYear{2018}
\acmDOI{XXXXXXX.XXXXXXX}
\acmConference[Conference acronym 'XX]{Make sure to enter the correct
  conference title from your rights confirmation email}{June 03--05,
  2018}{Woodstock, NY}
\acmISBN{978-1-4503-XXXX-X/2018/06}

\usepackage{xspace}
\usepackage[ruled,vlined,linesnumbered]{algorithm2e} 
\usepackage{listings}
\usepackage[skip=0pt]{subcaption}
\usepackage{caption}
\usepackage{enumitem}
\usepackage[most]{tcolorbox}
\usepackage{longtable}
\usepackage{array}
\usepackage{ragged2e}
\usepackage[dvipsnames]{xcolor}
\usepackage[ruled,vlined,linesnumbered]{algorithm2e} 
\usepackage{makecell}
\usepackage[table]{xcolor}
\usepackage{threeparttable}
\usepackage{booktabs, siunitx, array}
\usepackage{pifont}
\usepackage{bbm}
\usepackage{multirow}

\newcommand{\minihead}[1]{\vspace{0.5em}\noindent\textbf{#1}}

\newcommand{\eg}{e.g.}

\definecolor{forestgreen}{RGB}{10, 60, 10}

\newcommand{\myblock}{\scalebox{1.5}{$\bullet$}}
\newcommand{\myempty}{\scalebox{1.5}{$\circ$}}  

\newcommand{\rvbase}{\textsc{ReViSQL}\xspace}
\newcommand{\sn}{\textsc{ReViSQL-BIRD}\xspace}
\newcommand{\dn}{BIRD-Platinum\xspace}
\newcommand{\ts}{Text-to-SQL\xspace}

\begin{document}
\title{Human-Level Text-to-SQL via Reinforcement Learning on Verified Data, 
Without Pipeline Engineering}

\author{Yuxuan Zhu}
\affiliation{%
  \institution{University of Illinois}
}
\email{yxx404@illinois.edu}

\author{Tengjun Jin}
\affiliation{%
  \institution{University of Illinois}
}
\email{tengjun2@illinois.edu}

\author{Yoojin Choi}
\affiliation{%
  \institution{University of Illinois}
}
\email{yoojinc3@illinois.edu}

\author{Daniel Kang}
\affiliation{%
  \institution{University of Illinois\\Bridgewater AIA Labs}
}
\email{ddkang@illinois.edu}



\begin{abstract}
Translating natural language questions to SQL queries (\ts) is a long-standing 
problem in database research. Recent efforts have focused 
on improving \ts accuracy by building increasingly complex \emph{multi-stage 
large language model (LLM) pipelines}, layering task decomposition, schema 
linking, and model-based query selection on top of an LLM. Despite this growing 
pipeline complexity, a substantial gap (>10\%) between such systems and human 
experts persists on benchmarks, suggesting that pipeline engineering alone has 
hit a ceiling.

We show that human-level \ts performance is achievable by 
fine-tuning an LLM using reinforcement learning with verifiable rewards (RLVR) 
on clean data, without any pipeline components. In this paper, we identified the
dominant bottleneck for RLVR on \ts: existing training data contains pervasive 
annotation errors that mislead optimization. To address this, we developed 
a multi-round, expert-driven verification pipeline and used it to curate \dn, a 
dataset of 2.5k verified instances sampled from BIRD Train, with errors 
corrected in 61\% of instances. We show that fine-tuning Qwen3-235B-A22B on 
\dn yields consistent improvements (11--16\%) over BIRD Train on Arcwise-Plat (an
expert-verified version of BIRD Mini-Dev) and Spider2, outperforming
state-of-the-art open-source systems by 0.6--16\%. Furthermore, we diagnosed
two failure modes of standard RLVR on \ts. We find that (1) result-based rewards
have non-trivial false positive rates (32.8\% on average), and (2) models
systematically ignore the external knowledge in BIRD-style problems. To address 
these failure modes, we propose \sn, a specialized reward shaping method 
that combines result-based rewards with SQL equivalence verification and 
leverages process rewards for incentivizing external-knowledge use. We 
fine-tuned Kimi-K2.6 with \sn. On Arcwise-Plat, \sn-K2.6 is the first method to 
achieve human-level accuracy (92.96\%), outperforming the top five 
open-source systems on the leaderboard by 10--22\%.

\begin{figure}
    \centering
    \includegraphics[width=\linewidth]{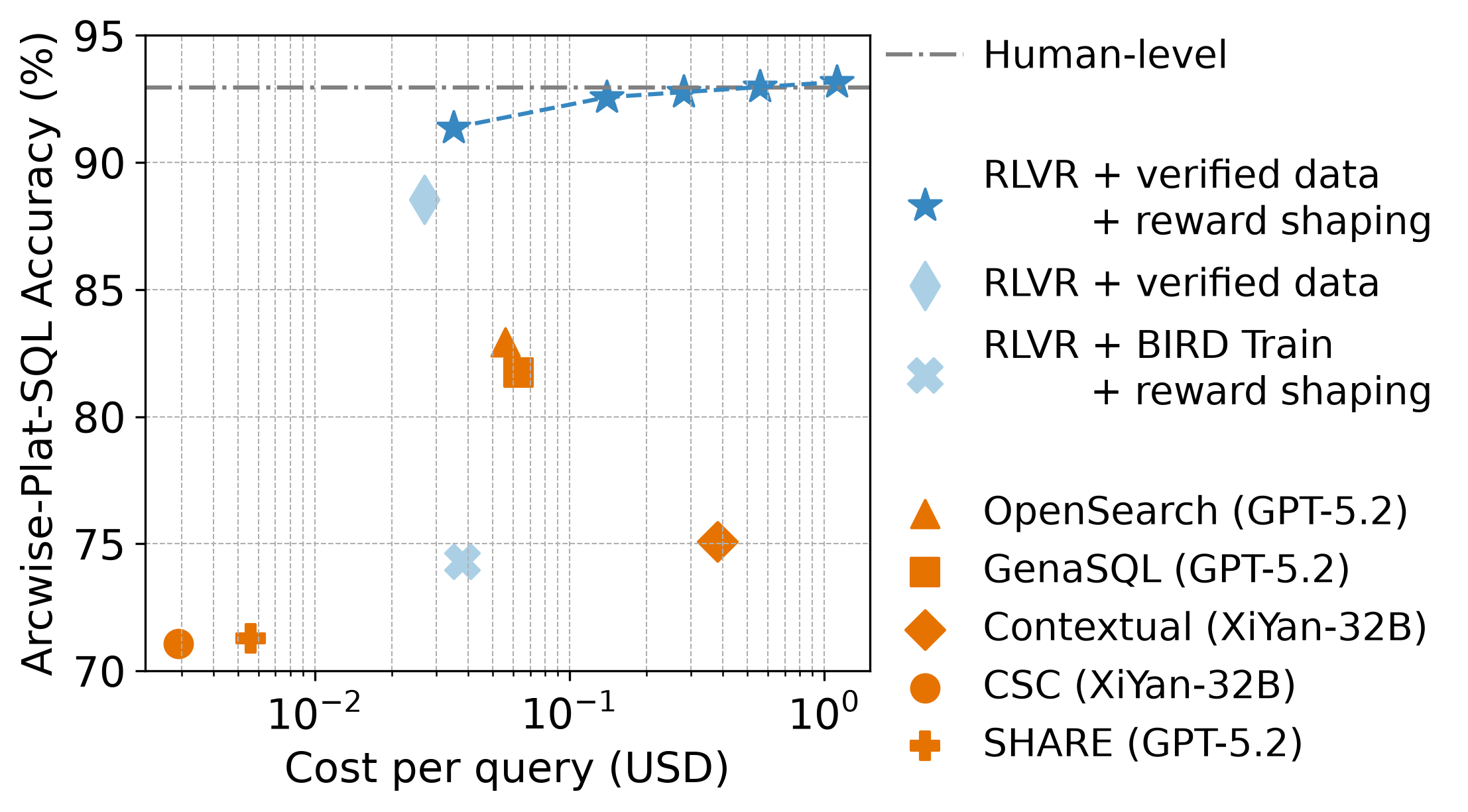}
    \caption{On Arcwise-Plat-SQL, an expert-verified BIRD Mini-Dev set 
    constructed by prior work \cite{jin2026pervasive,Arcwise-minidev}, \sn 
    achieves human-level performance \cite{bird-leaderboard} using RLVR 
    (Kimi-K2.6) with verified data and reward shaping. \sn obviates the need for 
    the complex pipelines \cite{xie2025opensearch,sheng2025csc,agrawal2025text2sql,
    donder2025cheaper,qu2025share} with 10--22\% higher accuracy.}
    \label{fig:cost-perf}
\end{figure}

\end{abstract}

\maketitle

\section{Introduction}

Translating natural language questions to SQL queries (\ts) is a foundational 
capability for democratizing data access \cite{li2014constructing,
androutsopoulos1995natural,sen2019natural,affolter2019comparative}. Recently, 
LLMs with strong reasoning and coding capabilities have emerged as its core 
engine \cite{katsogiannis2023survey,liu2025survey}. However, existing LLM-based 
methods still fall significantly short of human-level performance on 
the BIRD Leaderboard \cite{bird-leaderboard}, one of the most used benchmarks. 
Closing this gap is a precondition for deploying LLMs in production analytics, 
where SQL errors carry real cost~\cite{chen2025reliable}.

To improve the accuracy of out-of-the-box LLMs on \ts, existing work has built 
increasingly complex multi-stage pipelines that compensate for the failure modes 
of LLMs. These LLM pipelines layer task decomposition, schema linking, 
in-context example retrieval, and model-based candidate selection (Table 
\ref{tab:agents}) around the underlying LLM 
\cite{wang2025agentar,shkapenyuk2025automatic,
pourrezachase,liu2025xiyan,sheng2025csc,pourreza2025reasoning,
donder2025cheaper,xie2025opensearch,li2025omnisql,maamari2024death,
qu2025share,talaei2024chess,sheng2025slm,lialpha,li2024codes,
gao2024text,zhai2025excot,cohere2025command,li2024dawn}. 
Despite this growing complexity, the best such systems still 
underperform human experts by over 10\% on BIRD~\cite{bird-leaderboard}, 
suggesting that pipeline engineering alone has hit a ceiling.

We show that human-level \ts performance (92.96\% \cite{li2023can}) is achievable 
by fine-tuning an LLM with RLVR on clean data, without any of the pipeline 
components. We propose two training ingredients that together close the gap to 
human-level performance on \ts: a verification pipeline that produces clean
training data for RLVR (Section \ref{sec:data}), and reward shaping 
methods tailored to \ts problems (Section \ref{sec:method}). Our 
empirical evidence shows that clean training data is a dominant factor that 
brings RLVR within 3--4\% of human parity, while domain-specific reward shaping 
closes the residual gap by providing accurate learning signals. Once 
both are applied, the pipeline machinery built around the LLM becomes 
unnecessary for reaching human-level accuracy.

\minihead{\dn: a verified dataset for RLVR on \ts.}
Previous data-centric studies have identified pervasive annotation errors in 
existing \ts datasets, including BIRD and Spider2 
\cite{jin2026pervasive,wretblad2024understanding,liu2025nl2sql}, where over 50\% 
of their public instances contain errors. These errors lead to incorrect RLVR reward signals 
that cause fine-tuned models to make similar mistakes, limiting the 
effectiveness of prior RLVR efforts 
\cite{yao2025arctic,databricks-rlvr,ma2025sql}. To enable effective RLVR for 
\ts, we designed a verification pipeline involving 2--4 independent
manual correction rounds per data instance by SQL experts. Using this pipeline, 
we corrected 2.5k instances sampled from BIRD Train uniformly at random \cite{li2023can}. We 
identified and corrected annotation errors across 52\% of SQL queries, 26\% of 
natural language questions, and 18\% of external knowledge entries. We analyze 
the error patterns and release the guidelines of our verification pipeline in 
Section \ref{sec:data} for future \ts data curation.

\minihead{\rvbase: RLVR on verified data.} We introduce \rvbase, a recipe that 
combines RLVR with \dn. Despite its simplicity, \rvbase significantly improves  
accuracy and achieves generalizability across benchmarks. We evaluated on 
two verified versions \cite{jin2026pervasive} of BIRD Mini-Dev, 
Arcwise-Plat-Full (errors in questions, knowledge, and gold SQL fixed) 
and Arcwise-Plat-SQL (only gold SQL fixed; noisy questions and knowledge 
preserved). Fine-tuning Qwen3-235B-A22B with \rvbase (\rvbase-235B) yields 11\% 
and 16\% higher accuracy than RLVR with BIRD Train on Arcwise-Plat-Full and -SQL. 
Furthermore, we evaluated on Spider2-SQLite and 
Spider2-Snow \cite{lei2024spider}, where \rvbase-235B yields 12\% and 14\% 
higher accuracy than RLVR on BIRD Train. Beyond demonstrating the value of 
verified training data, \rvbase-235B also matches or outperforms prior 
open-source pipelines at lower inference costs. Compared to the best
open-source LLM pipelines (Table \ref{tab:spider2sqlite}), \rvbase-235B 
achieves 4.4\%, 0.6\%, and 2.2\% higher greedy-decoding accuracy on 
Arcwise-Plat-SQL, -Full, and Spider2-SQLite, with 7--8$\times$ 
lower costs. For Spider2-Snow, \rvbase-235B with 
self-consistency\footnote{We use the standard self-consistency procedure for 
inference-time scaling~\cite{wang2022self,dong2023c3,gao2023text,
xie2025opensearch,sheng2025csc}. See Section~\ref{sec:eval-headline} for full 
specification.} 
over 128 candidates outperforms open-source LLM agents that use larger models such 
as DeepSeek-V3 and Qwen3-Coder.

\minihead{\sn: achieving human parity on BIRD.} Although \rvbase significantly 
improves RLVR, fine-tuning one of the best open-source LLMs, Kimi-K2.6, 
using \rvbase still underperforms human parity by 3--4\% on Arcwise-Plat 
datasets. To close the residual gap, we diagnosed two failure modes of 
\rvbase and developed a reward shaping method, \sn, to address them 
(Section~\ref{sec:method}). First, we find that the result-based reward, the 
widely used reward in \ts RLVR \cite{yao2025arctic,databricks-rlvr,ma2025sql}, 
leads to an empirical false positive rate of 32.8\%. This is because a 
semantically incorrect generated SQL query can yield the same results as the 
gold query on a concrete database. We mitigate such false positive rewards by 
integrating verification-based grading \cite{he2024verieql} with result-based 
grading \cite{liu2025skyrlsql,ma2025sql}. Second, we identify that fine-tuned 
models systematically ignore the external-knowledge entries provided in BIRD 
instances, defaulting to their pretraining priors over domain semantics. To 
address this failure, we incentivize the model to effectively leverage the 
external knowledge provided in a \ts problem by assigning rule-based process 
rewards in addition to the widely used outcome rewards 
(Section \ref{subsec:shaping}). 

We fine-tuned Kimi-K2.6 with \sn on \dn and evaluated the resulting model,
\sn-K2.6, on Arcwise-Plat-Full and Arcwise-Plat-SQL \cite{jin2026pervasive}. 
With greedy decoding, \sn-K2.6 achieves 93.8\% accuracy on Arcwise-Plat-Full and 91.4\% accuracy 
on Arcwise-Plat-SQL. With self-consistency over 16 candidates, \sn-K2.6 reaches 
94.2\% and 93.0\% respectively. On both Arcwise-Plat benchmarks, our 
pipeline-free results match the proxy human-level performance (92.96\%) measured 
on BIRD Test \cite{li2023can}. Furthermore, \sn-K2.6 outperforms OpenSearch 
\cite{xie2025opensearch}, the strongest prior open-source pipeline powered by 
GPT-5.2, by 8.4\% on Arcwise-Plat-SQL and 5.6\% on Arcwise-Plat-Full.

\vspace{0.5em}
We summarize our contributions as follows:
\vspace{-0.35em}
\begin{enumerate}[leftmargin=*]
    \item \textbf{A paradigm shift for \ts}. To our knowledge, we are the first 
    to demonstrate that human-level \ts is achievable through clean data 
    and RLVR alone, without multi-stage pipelines, suggesting a paradigm shift
    from system design to high-quality data curation for \ts.
    \item \textbf{\dn and \rvbase}. We curated \dn, a verified dataset 
    with 2.5k high-quality \ts instances. We show that fine-tuning LLMs with 
    \rvbase (RLVR on \dn) achieves consistent improvements on Arcwise-Plat 
    (verified BIRD Mini-Dev) \cite{jin2026pervasive} and Spider2 
    \cite{lei2024spider}, matching or outperforming state-of-the-art LLM 
    pipelines with lower costs.
    \item \textbf{\sn}. We diagnosed two failure modes of RLVR on \ts: 
    false-positive result-based rewards and systematic neglect of external 
    knowledge. To address them, we propose \sn, a domain-specialized reward  
    shaping method leveraging verification-aware rewards and process rewards. 
    We show that fine-tuning Kimi-K2.6 using \sn achieves human-level 
    performance on Arcwise-Plat for the first time.
\end{enumerate}
\section{Related Work}

In this section, we review the landscape of existing \ts methods 
(Section~\ref{sec:bg-method}) and examine the critical limitations of current 
\ts datasets (Section~\ref{sec:bg-data}).

\subsection{Text-to-SQL Methods} \label{sec:bg-method}
We categorize prior work into three distinct paradigms: multi-stage LLM 
pipelines, training with massive synthetic data, and RLVR.

\begin{table}[t]
    \centering
    \footnotesize
    \caption{Comparison of the pipeline complexity among top-30 BIRD Leaderboard 
    methods that have disclosed technical details. Each filled dot marks a 
    component the method requires at inference in addition to SQL query 
    generation. \sn exceeds these methods on accuracy while requiring none 
    of them.}
    \label{tab:agents}

    \rowcolors{2}{gray!8}{white}
    \begin{tabular}{l|c c c c c}
    \toprule
    Method & \makecell{Dynamic\\pipeline} & \makecell{Dynamic\\prompts} & \makecell{Schema\\linking} & \makecell{Iterative\\revision} & \makecell{Selection} \\ 
    \midrule
    JoyDataAgent \cite{JoyAgent-JDGenie}    & \myblock & \myblock & \myblock & \myblock & \myempty \\ 
    SiriusAI \cite{jiang2024siriusbi}       & \myblock & \myempty & \myblock & \myblock & \myempty \\
    Agentar-Scale \cite{wang2025agentar}    & \myempty & \myblock & \myempty & \myblock & \myblock \\ 
    SHARE \cite{qu2025share}                & \myempty & \myblock & \myblock & \myblock & \myempty \\ 
    OpenSearch \cite{xie2025opensearch}     & \myempty & \myblock & \myblock & \myblock & \myempty \\ 
    Distillery \cite{maamari2024death}      & \myempty & \myblock & \myblock & \myblock & \myempty \\ 
    CHASE-SQL \cite{pourrezachase}          & \myempty & \myempty & \myblock & \myblock & \myblock \\ 
    ReasoningSQL \cite{pourreza2025reasoning}  &  \myempty & \myempty & \myblock & \myblock & \myblock \\ 
    AskData \cite{shkapenyuk2025automatic}  & \myempty & \myblock & \myblock & \myempty & \myempty \\ 
    GenaSQL \cite{donder2025cheaper}        & \myempty & \myempty & \myblock & \myempty & \myblock \\ 
    XiYan-SQL \cite{liu2025xiyan}           & \myempty & \myempty & \myblock & \myempty & \myblock \\ 
    CSC-SQL \cite{sheng2025csc}             & \myempty & \myempty & \myempty & \myempty & \myblock \\ 
    Contextual \cite{agrawal2025text2sql}   & \myempty & \myempty & \myempty & \myempty & \myblock \\ 
    \hline \specialrule{0.5pt}{2pt}{2pt}
    \textbf{\sn}                     & \myempty & \myempty & \myempty & \myempty & \myempty \\
    \bottomrule
    \end{tabular}
\end{table}

\minihead{Multi-stage LLM pipelines.}
Advancements in \ts, particularly on benchmarks including BIRD 
\cite{li2023can} and Spider2 \cite{lei2024spider}, have been heavily driven by 
the development of complex multi-stage LLM pipelines \cite{wang2025agentar,
shkapenyuk2025automatic,pourrezachase,liu2025xiyan,sheng2025csc,
donder2025cheaper,xie2025opensearch,qu2025share,talaei2024chess,sheng2025slm,
lialpha,li2024codes}. As shown in Table~\ref{tab:agents}, top-ranking methods 
on BIRD rely on hand-engineered auxiliary modules around the LLM at inference 
time. We summarize prior work by the failure mode each component addresses, and 
argue that RLVR fine-tuning targets the same failure modes directly in the model, 
rather than working around them externally.

General-purpose LLMs struggle to map natural-language entities to the right
tables and columns when the data schema is complex, leading to invalid column 
references and incorrect joins. Schema linking and dynamic prompts patch this 
externally and are widely used in prior systems, including 
AskData~\cite{shkapenyuk2025automatic}, SHARE~\cite{qu2025share}, 
OpenSearch~\cite{xie2025opensearch}, and Distillery~\cite{maamari2024death}. 
They pre-filter the schema, rewrite questions, and retrieve few-shot examples to 
narrow the LLM's search space before generation. RLVR addresses the same failure 
mode internally. By rewarding the exploration trajectories that lead to correct
queries, RLVR teaches the model to explore the database and ground its outputs 
in the schema, eliminating the need to externally link a question to the schema.

Furthermore, LLMs frequently produce queries that fail to execute or return 
incorrect results on the first attempt due to syntax errors, type mismatches, or 
subtle semantic mistakes. Iterative revision loops patch this externally. Prior 
methods, such as JoyDataAgent-SQL~\cite{JoyAgent-JDGenie}, 
Agentar-Scale-SQL~\cite{wang2025agentar}, and CHASE-SQL~\cite{pourrezachase}, 
execute candidate queries against the target database and feed execution errors 
back to the LLM for iterative debugging. RLVR addresses the same failure mode by
penalizing incorrect queries that would trigger revision at inference, pushing 
the model toward queries that execute correctly the first time.

Even with refinement, individual generations remain unreliable. Hence, many 
systems trade compute for accuracy via model-based, post-generation selection.
GenaSQL~\cite{donder2025cheaper}, XiYan-SQL~\cite{liu2025xiyan}, and
CSC-SQL~\cite{sheng2025csc} sample multiple candidates and use secondary reward 
models to score, filter, or vote among them. 
JoyDataAgent-SQL~\cite{JoyAgent-JDGenie} further uses an LLM to plan and
route across this multi-stage pipeline. These methods not only add inference 
complexity but also incur extra cost to train reward models. RLVR addresses this 
failure mode by shifting the distribution of generations themselves. For example,
GRPO \cite{shao2024deepseekmath}, a typical RLVR objective, maximizes the 
expected accuracy of a group of generations per question, training a model to 
generate correct queries with fewer candidates. Empirically, we show that, on a 
verified benchmark without noisy questions (Arcwise-Plat-Full) \cite{jin2026pervasive}, 
greedy decoding from an RLVR model matches human-level accuracy without selection 
mechanisms. On a benchmark with noisy questions (Arcwise-Plat-SQL), an RLVR 
model achieves human parity using model-free, self-consistency generation
with 16 candidates per question (Section \ref{sec:eval-headline}).

While each module provides distinct utility in patching a specific LLM
failure mode, their composition introduces inference-time overhead and cascading 
error propagation~\cite{wang2025agentar,cemri2025multi}. For example, a wrong 
schema-linking result corrupts every downstream stage. More fundamentally, these 
pipelines remain limited by the underlying model's intrinsic SQL capability. \sn 
instead trains the model itself, using RLVR on clean data with tailored 
reward shaping, eliminating the external machinery that exists to compensate for
gaps in the underlying model's reasoning.

\minihead{Training on synthetic data.}
In parallel to building LLM pipelines, prior work has also explored 
data-centric approaches, developing large-scale synthetic \ts data for training 
LLMs \cite{li2025omnisql,pourreza2024sql,wolff2025sqale}. SynSQL-2.5M 
\cite{li2025omnisql} developed a pipeline to generate 2.5 million synthetic \ts 
instances, aiming to cover every conceivable SQL pattern. SQL-GEN 
\cite{pourreza2024sql} addresses the specific challenge of dialect adaptation, 
covering common dialects including PostgreSQL, BigQuery, and SQLite. SQaLe 
\cite{wolff2025sqale} takes a different angle, focusing on schema realism. It 
generates queries based on 135,875 real-world schemas, exposing models to 
``messy'' production databases (\eg, thousands of tables, obscure column names).

While synthetic data enables large-scale training 
\cite{li2025omnisql,pourreza2024sql} and improves robustness 
\cite{wolff2025sqale}, it fails to address the inherent gap between human data 
and synthetic logic. These datasets typically generate natural language 
questions with clear, consistent logic, which does not reflect real-world 
deployment. Our research suggests that rigorous curation is more important than 
large-scale generation.

\minihead{RLVR for \ts.}
RLVR has emerged as a promising alternative to multi-stage pipelines for \ts,
shifting investment from inference-time orchestration to training-time 
optimization of the underlying LLM. Prior work has applied RLVR to \ts tasks 
\cite{yao2025arctic,ma2025sql,infly-sql,papicchio2025think2sql,pourreza2025reasoning}. 
For example, Arctic-Text2SQL-R1~\cite{yao2025arctic} fine-tuned open-weight 
models using result-based rewards on BIRD Train. Concurrent work from
Databricks~\cite{databricks-rlvr} and SQL-R1~\cite{ma2025sql} adopted similar
execution-grading approaches. Moreover, ReasoningSQL~\cite{pourreza2025reasoning}
extended this with partial reward terms targeting approximate matching of SQL 
queries. 

However, prior RLVR efforts still fail to outperform complex LLM pipelines 
\cite{bird-leaderboard}. We argue this stems from several limitations that no 
prior work has addressed. First, all of these efforts train, at least in part, 
on the official BIRD Train set whose data quality has not yet been fully verified. 
Second, prior efforts rely on standard result-based grading as the reward signal, 
which we show in Section~\ref{sec:method} is unreliable on \ts. In addition, 
domain-specific problems include external knowledge that the model must 
integrate (\eg, evidence entries in BIRD \cite{li2023can} and external knowledge 
in Spider2 \cite{lei2024spider}), but standard outcome-based rewards 
provide no signal about whether the model actually used it. Finally, prior work 
establishes the effectiveness of RLVR only for relatively small base models 
(7B--32B) evaluated on unverified benchmarks. In this work, we 
address these limitations simultaneously (Section \ref{sec:data},  
\ref{sec:method}, and \ref{sec:eval}).

\subsection{Noise in Text-to-SQL Datasets} \label{sec:bg-data}
Pervasive annotation errors hinder the rigorous evaluation and training of 
\ts models. In BIRD \cite{li2023can} and Spider2 \cite{lei2024spider}, 
two widely used datasets, recent audits have uncovered errors in up to 52\% of 
questions, SQL queries, and external knowledge across both evaluation and 
training sets~\cite{jin2026pervasive,Arcwise-minidev,wretblad2024understanding,
pourreza2023evaluating,liu2025nl2sql}, causing concerns in both the development 
and evaluation of \ts methods.

\minihead{Errors in evaluation sets.}
The annotation errors in the BIRD Dev set have been heavily investigated.
\citet{jin2026pervasive} identified annotation errors in 52\% of instances. 
Similarly, \citet{wretblad2024understanding} analyzed the financial domain 
subset, finding that ambiguous questions and incorrect gold SQL queries 
introduced noise into 15--49\% of instances per database. While 
\citet{liu2025nl2sql} proposed labeling these errors to test error-detection 
capabilities, they did not correct them. To enable reliable evaluation, 
\citet{Arcwise-minidev} and subsequently \citet{jin2026pervasive} released two 
verified variants of the BIRD Mini-Dev that we used as our evaluation benchmarks: 
Arcwise-Plat-Full and Arcwise-Plat-SQL. These two benchmarks differ in which
sources of error are corrected (Table \ref{tab:arcwise};
Appendix~\ref{appendix:arcwise-case-study} walks through a representative
instance). Arcwise-Plat-Full delivers a fully cleaned version with errors 
corrected in all three sources: 26.5\% of gold SQL queries, 17.7\% of natural 
language questions, and 19.4\% of external knowledge entries. Arcwise-Plat-SQL 
corrects only the gold SQL queries (16.1\% of instances), while 
\emph{deliberately preserving} the factual errors and ambiguities in questions 
and external knowledge~\cite{jin2026pervasive}. The two benchmarks thus evaluate 
different capabilities. Arcwise-Plat-Full measures accuracy under clean inputs, 
while Arcwise-Plat-SQL measures robustness to the noisy questions and external 
knowledge that are closer to real-world \ts deployments \cite{liu2025survey}.

\begin{table}
    \centering
    \caption{Correction statistics of two expert-verified datasets based on 
    BIRD mini-Dev.}
    \label{tab:arcwise}
    \begin{tabular}{ccc}
        \toprule
        Correction  & Arcwise-Plat-Full & Arcwise-Plat-SQL \\
        \midrule
        SQL query    & 132 (26.5\%)      & 80 (16.1\%) \\
        Question    & 88  (17.7\%)      & 0  \\
        External knowledge & 97 (19.4\%)& 0  \\
        \bottomrule
    \end{tabular}
\end{table}

\minihead{Errors in training sets.}
While prior work has verified the BIRD Mini-Dev evaluation set, \ts training data
remains largely unverified. \citet{pourreza2023evaluating} partially categorized
errors in the training set into three types: sorting failures (ties in 
\texttt{ORDER BY}), schema matching ambiguity, and incorrect content assumptions. 
They detected errors in 18.2\% of a subset of problems, which likely 
underestimates the true noise level. To date, there are no large-scale, publicly 
available corrections for the BIRD Train set. This lack of clean training data 
represents a fundamental barrier to training models via reinforcement learning.
\section{Curating \dn: A Verified Dataset for Text-to-SQL RLVR}
\label{sec:data}

\begin{figure*}
    \centering
    \includegraphics[width=\textwidth]{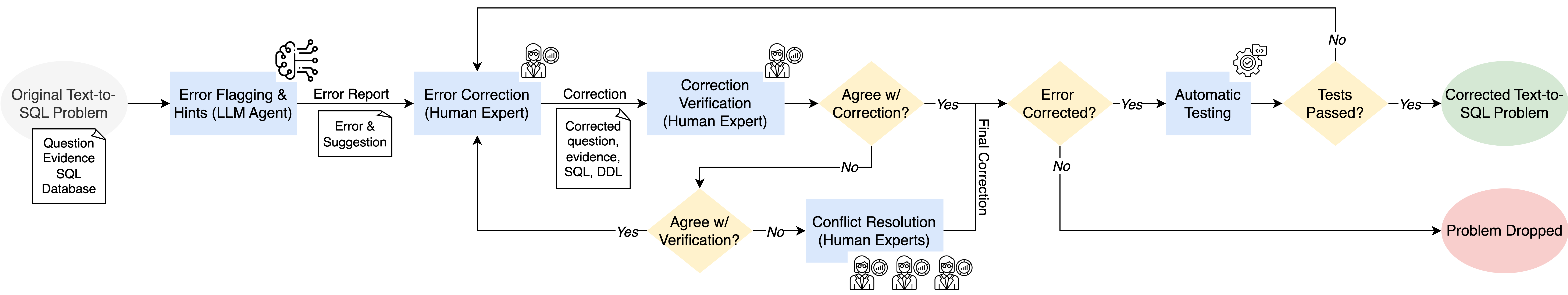}
    \caption{Our hybrid LLM--human-expert pipeline for correcting text-to-SQL 
    annotation errors in BIRD.}
    \label{fig:correction-pipeline}
\end{figure*}

\lstset{
    language = SQL,
    keywordstyle = {\ttfamily \color{blue}},
    basicstyle=\ttfamily\footnotesize,
    breaklines = true,
    breakatwhitespace = true,
    keepspaces=true, 
    morekeywords={with}
}

We constructed \dn, a verified dataset of 2.5k instances sampled from BIRD Train 
and corrected through a hybrid LLM--human-expert pipeline. In this section, 
we describe the design principles behind our correction process 
(\S\ref{subsec:checklist}), the five-stage pipeline that implements them 
(\S\ref{subsec:pipeline}), a taxonomy of the errors we encountered with worked 
examples (\S\ref{subsec:errors}), and summary statistics for both the 
construction process and the resulting dataset (\S\ref{subsec:con-stats}).

\subsection{Design Principles and Verification Checklist} 
\label{subsec:checklist}

\minihead{Soundness over completeness.} 
For a verified dataset to improve RLVR training, two properties of data 
cleaning are essential.
\begin{enumerate}[leftmargin=*]
    \item Every correction must target a \emph{genuine} annotation error.
    \item Corrections themselves must not introduce new errors. 
\end{enumerate}
We optimize the pipeline for these two soundness properties, even at the cost
of completeness. We do not claim to find every error in every instance. Instead, 
we ensure that every change we make is principled, defensible, and verified by an 
independent expert.

\minihead{Checklist.} 
To achieve soundness, we develop a checklist that human experts apply to every 
instance. The checklist enumerates the specific failure modes we observed during 
pilot annotation and serves as both a guide for correction and verification. We 
organize it around the components of a \ts instance: the natural language 
question, the external knowledge, and the gold SQL query.

\vspace{0.4em}
\noindent For each natural language question, we check:
\begin{enumerate}[leftmargin=*]
\item Is the question answerable from the given database?
\item Does the question admit a single interpretation?
\item Is the expected output, including whether duplicates should be removed, 
unambiguously specified?
\item Does the question yield a non-empty answer?
\item Do all filter conditions in the question constrain the result?
\end{enumerate}
For each external knowledge entry, we check:
\begin{enumerate}[leftmargin=*]
\item Is the external knowledge relevant to answering the question?
\item Is the external knowledge consistent with the question?
\end{enumerate}
For each gold SQL query, we check:
\begin{enumerate}[leftmargin=*]
\item Does the SQL query output all and only the required columns?
\item Does the SQL query handle ties correctly?
\item Does the SQL query handle duplicates correctly?
\item Does the SQL query handle missing values correctly?
\item Does the SQL query use the correct tables via appropriate joins and set 
operations?
\item Are all filter conditions from the question correctly translated?
\item Are aggregations computed over the correct partitions?
\item Does string-value extraction generalize across the full database, not just 
the sampled rows?
\end{enumerate}

\subsection{Construction Pipeline} \label{subsec:pipeline}
\label{sec:data-curation}

During data construction, each instance passes through five stages designed to 
combine the breadth of LLM-based error detection with the precision of human 
expert judgment (Figure~\ref{fig:correction-pipeline}). The pipeline 
incorporates independent correction with independent verification to avoid 
propagating mistakes of an expert to the final dataset.

\minihead{Stage 1: LLM-assisted error flagging.} 
We applied an LLM reviewer \cite{jin2026pervasive} based on OpenAI's o3 to detect
annotation errors and provide correction suggestions. The agent can execute
SQL queries on the example database via function calls, with a budget of 30
executions per instance. The agent generates an error report describing errors
in the question, external knowledge, and SQL query, as well as a suggested
correction of the SQL query.

\minihead{Stage 2: Independent expert correction.} 
A human expert independently investigates the instance using the checklist 
(\S\ref{subsec:checklist}), \emph{before} reading the LLM agent's report. The 
order ensures the expert detects errors the LLM missed rather than completely 
relying on the LLM's findings. The expert then incorporates the LLM report and 
produces a draft correction covering the question, external knowledge, and SQL 
query.

\minihead{Stage 3: Independent verification.} 
A second expert independently re-investigates the instance and the draft 
correction, applying the same checklist. The verifier confirms each correction 
is sound and searches for errors the first expert may have missed. Disagreement 
between the correction and verification experts triggers revision by a third 
expert. If the third expert disagrees with the verification result, we proceed to 
conflict resolution.

\minihead{Stage 4: Conflict resolution.} 
When correction and verification experts disagree on the substance of a 
correction, three experts convene to reach consensus. In our process, every 
conflict was resolved at this stage. The one outcome that we treat specially is
when the experts agree that no correction is possible. This is typically because
the question cannot be answered from the given database 
(\eg, \S\ref{subsec:errors-question}, item~1). In these cases, we drop the 
instance.

\minihead{Stage 5: Automatic testing.} 
We then run the following two automated checks on the corrected instance. If either check fails, 
we route the instance back to manual revision and verification.
\begin{enumerate}[leftmargin=*]
\item The corrected SQL query outputs a non-empty result.
\item The output of the corrected SQL query must differ from the output of the original SQL query.
\end{enumerate}

\subsection{Taxonomy of Annotation Errors}
\label{subsec:errors}

The corrections produced by our pipeline form a more comprehensive taxonomy of
annotation errors than those investigated in prior work \cite{jin2026pervasive,liu2025nl2sql,wretblad2024understanding}. We describe the taxonomy to explain 
what \dn corrects and hopefully to inspire future \ts data curation. We 
group errors into four categories: question and external knowledge errors,
gold SQL query errors, schema errors, and grading-method errors.

\subsubsection{Question and External Knowledge Correction} 
\label{subsec:errors-question}

We identified and corrected the following three types of errors in the natural
language questions and evidence:
\begin{enumerate}[leftmargin=*]

\item \underline{Unanswerable from schema.} Some questions cannot be answered
from the provided database. We dropped such instances. In the following example, 
the question presumes a sale-status attribute that the database does not
contain:
\begin{tcolorbox}[
        colback=gray!5!white, 
        colframe=gray!20!white, 
        arc=4pt, 
        boxrule=1pt, 
        left=3pt, right=3pt, top=3pt, bottom=3pt,
        breakable
    ]
\footnotesize
\textit{Original Question}: Among the products, how many of them were \textcolor{red}{discontinued} in production?

\vspace{0.5em}
\textit{Action}: (Instance dropped)
\end{tcolorbox}

\item \underline{Empty result.} We should not use filter conditions that yield 
empty answer sets as RLVR targets, because a model can achieve maximum reward on these instances
with trivial queries such as \texttt{SELECT 1 WHERE False}, a classic
reward-hacking pathology \cite{skalse2022defining}. We revise filter conditions
to ensure non-empty results.

\item \underline{External knowledge-question mismatch.} When external 
knowledge entries contradict a question, we align the external knowledge to the
question. For example, we found a question asking for ``at least 200 movies'' 
that has external knowledge ``> 200''. We then updated the external
knowledge to ``>=200''.
\end{enumerate}

\subsubsection{Gold SQL Query Errors} 
\label{subsec:errors-sql}

\begin{enumerate}[leftmargin=*]
\item \underline{Mishandled missing data.} In practice, \texttt{NULL} values, 
sentinel values, or indicator columns often mark missing data. Aggregation or 
subsequent data operations should exclude them to produce meaningful results when
necessary. In the following example, missing values in \texttt{weight} and 
\texttt{height} corrupt the \texttt{MAX} and \texttt{MIN} aggregations of BMI:
\begin{tcolorbox}[
        colback=gray!5!white, 
        colframe=gray!20!white, 
        arc=4pt, 
        boxrule=1pt, 
        left=3pt, right=3pt, top=3pt, bottom=3pt,breakable
    ]
    \footnotesize
    \textbf{Question:} What are the basketball players' BMI ranges? \\[1ex]
    \textbf{Original SQL:} 
\begin{lstlisting}
SELECT
    MIN(weight * 1.0 / (height * height) * 703),
    MAX(weight * 1.0 / (height * height) * 703)
FROM players
\end{lstlisting}
    \textbf{Corrected SQL:} 
\begin{lstlisting}
SELECT ... FROM ... -- same as the original
WHERE weight > 0 AND height > 0;
\end{lstlisting}
    \end{tcolorbox}

\item \underline{Unhandled ties.} When a question asks for an extreme value, 
\texttt{ORDER BY ... LIMIT 1} silently discards tied rows. We corrected such 
queries to return all rows matching the extreme value:
\begin{tcolorbox}[
        colback=gray!5!white, 
        colframe=gray!20!white, 
        arc=4pt, 
        boxrule=1pt, 
        left=3pt, right=3pt, top=3pt, bottom=3pt,
        breakable
    ]
    \footnotesize
    \textit{Question}: Name the movies with the most ratings.

    \vspace{0.5em}
    \textit{Original SQL Query}: 
\begin{lstlisting}
SELECT movie_title FROM movies 
GROUP BY movie_title 
ORDER BY COUNT(movie_title) DESC LIMIT 1
\end{lstlisting}

    \vspace{0.5em}
    \textit{Corrected SQL Query}: 
\begin{lstlisting}
WITH cte AS (
    SELECT m.movie_title title, count(*) cnt
    FROM movies AS m JOIN ratings AS r 
    ON m.movie_id = r.movie_id 
    GROUP BY m.movie_id, m.movie_title )
SELECT title FROM cte
WHERE cnt = (SELECT MAX(cnt) FROM cte)
\end{lstlisting}
    \end{tcolorbox}

\item \underline{Wrong or missing joins.} The example above also illustrates 
this category. The original SQL query miscounts movies without ratings, while the 
corrected query joins \texttt{movies} and \texttt{ratings}.

\item \underline{Wrong or missing predicates.} In the example below, the 
question specifies ``over 1,000'', while the original query uses 
``> 10'':
\begin{tcolorbox}[
        colback=gray!5!white, 
        colframe=gray!20!white, 
        arc=4pt, 
        boxrule=1pt, 
        left=3pt, right=3pt, top=3pt, bottom=3pt,
        breakable
    ]
    \footnotesize
    \textit{Question}: Among all the users that have posted a tweet with over 1000 likes, how many of them are male?
\lstset{escapeinside={<@}{@>}}

    \vspace{0.5em}
    \textit{Original SQL Query}: 
\begin{lstlisting}
SELECT COUNT(T1.TweetID) 
FROM twitter AS T1 INNER JOIN user AS T2 
ON T1.UserID = T2.UserID 
WHERE T1.Likes <@\textcolor{red}{> 10}@> AND T2.Gender = 'Male'
\end{lstlisting}

    \vspace{0.5em}
    \textit{Corrected SQL Query}: 
\begin{lstlisting}
SELECT COUNT(DISTINCT T1.UserID) 
FROM ... JOIN ... -- same as the original
WHERE T1.Likes <@\textcolor{teal}{> 1000}@> AND T2.Gender = 'Male'
\end{lstlisting}
    \end{tcolorbox}

\item \underline{Unhandled duplicates.} The example above also illustrates the 
failure to handle duplicated values when aggregating. After joining 
\texttt{twitter} and \texttt{user}, the column \texttt{UserID} is no longer 
unique. Hence, counting users requires \texttt{DISTINCT}. In our experiments, 
this error leads to significant consequences for RLVR. In an ablation, a model 
trained on uncorrected data applied \texttt{DISTINCT} correctly in only 5 of the
30 problems that require it, indicating that the model picked up the failure 
modes in the training data.

\item \underline{Wrong or missing output columns.} We corrected queries whose 
\texttt{SELECT} clauses omit columns requested by the question or include 
columns the question does not request.

\item \underline{Wrong or missing group keys.} We corrected \texttt{GROUP BY} 
clauses that aggregate over incorrect partitions. 

\item \underline{Wrong or missing tables.} We added tables that the original query
failed to join or set-combine, using the data schema to determine the correct
relation.

\item \underline{Wrong ranking criteria.} When \texttt{ORDER BY} ranks by 
expression inconsistent with the question, we corrected it:
\begin{tcolorbox}[
        colback=gray!5!white, 
        colframe=gray!20!white, 
        arc=4pt, 
        boxrule=1pt, 
        left=3pt, right=3pt, top=3pt, bottom=3pt,
        breakable
    ]
    \footnotesize
    \textit{Question}: List the product ID of the top five products, by descending order, the number of quantities ordered.

    \vspace{0.5em}
    \textit{Original SQL Query}:
    \begin{lstlisting}
SELECT ProductID FROM Products 
ORDER BY UnitsInStock DESC LIMIT 5
\end{lstlisting}

    \vspace{0.5em}
    \textit{Corrected SQL Query}: 
    \begin{lstlisting}
SELECT `ProductID` FROM `OrderDetails` 
GROUP BY `ProductID` 
ORDER BY SUM(`Quantity`) DESC LIMIT 5
\end{lstlisting}
    \end{tcolorbox}

\item \underline{Execution Errors}: We fixed syntax and reference errors that 
prevent the gold query from executing. The example above also demonstrates this.
The original query references a column \texttt{UnitsInStock} that does not exist
in the table \texttt{Products}.
\end{enumerate}

\begin{table*}
    \centering
    \caption{Structural complexity comparison between the original BIRD Train
    and \dn. Values for the datasets denote the average count of each structural
    component per query; the difference is the relative change from BIRD Train to
    \dn. We find that gold SQL queries exhibit a significant increase in complexity
    across every structural metric.}
    \label{tab:data-stats}
    \begin{threeparttable}
    \begin{tabular}{c|cccccccc}
        \toprule
        Per-query stats & \# tables & \# joins & \# functions & \# aggregations & \# set operations & \# subqueries & \# CTEs & \# window func. \\
        \midrule
        BIRD Train & 2.08 & 1.02 & 1.67 & 0.62 & 0.0024 & 0.088 & 0.0 & 0.0012 \\
        \dn        & 2.36 & 1.06 & 2.10 & 0.80 & 0.024 & 0.26 & 0.11 & 0.0012 \\
        Difference & +13.5\% & +3.9\% & +25.7\% & +29.0\% & +900\% & +195.5\% & ---\tnote{$*$} & +2.0\% \\
        \bottomrule
    \end{tabular}
    \begin{tablenotes}
        \footnotesize
        \item[] Differences are computed from unrounded values.
        \item[$*$] The original BIRD Train contains no common table expressions, so
        the relative change is undefined; \dn introduces 0.11 CTEs per query.
    \end{tablenotes}
    \end{threeparttable}
\end{table*}

\begin{figure*}[t!]
    \begin{subfigure}{0.32\textwidth}
        \includegraphics[width=\linewidth]{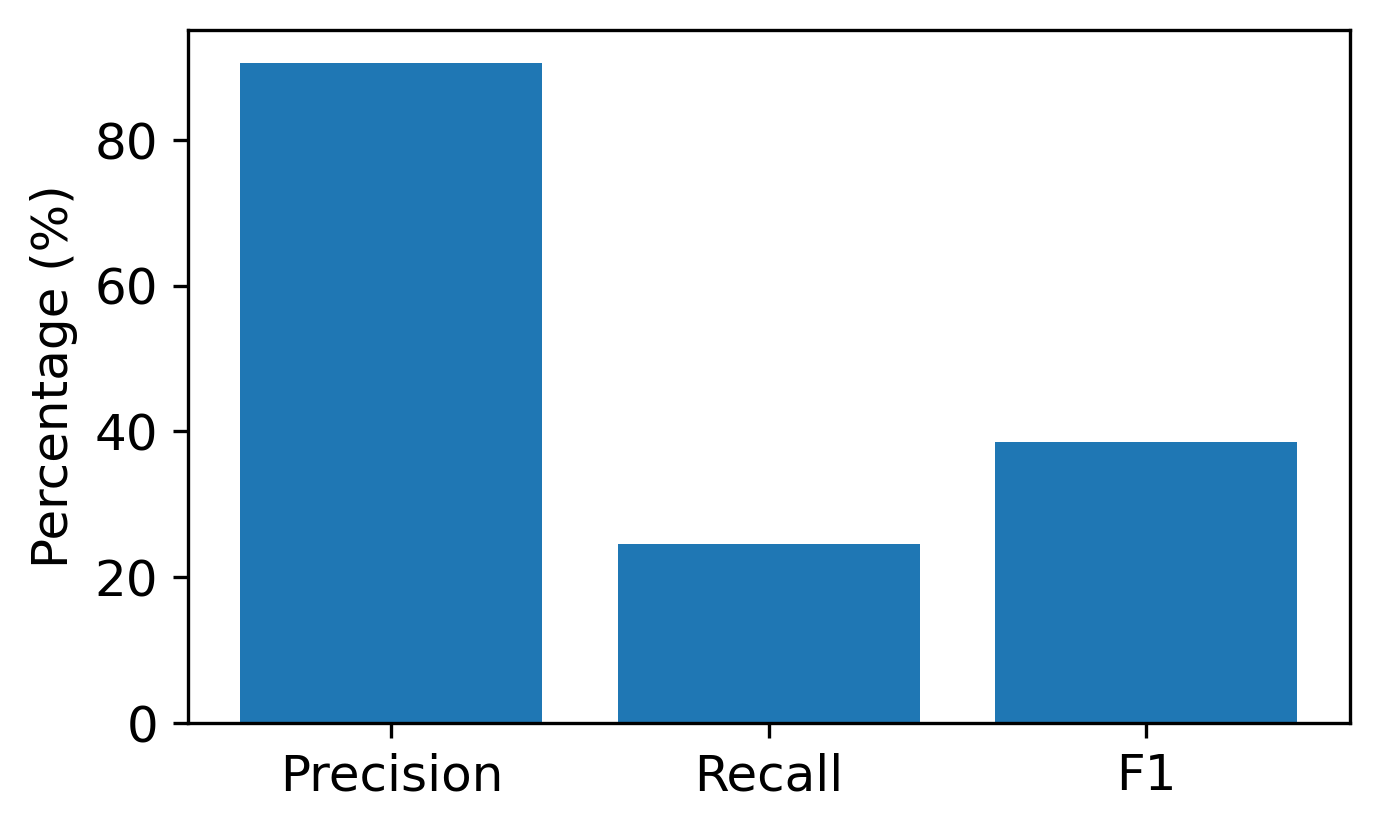}
        \caption{LLM reviewer demonstrates high precision but critically low 
        recall.} 
        \label{fig:llm-reviewer-metrics}
    \end{subfigure}\hspace*{\fill}
    \begin{subfigure}{0.32\textwidth}
        \includegraphics[width=\linewidth]{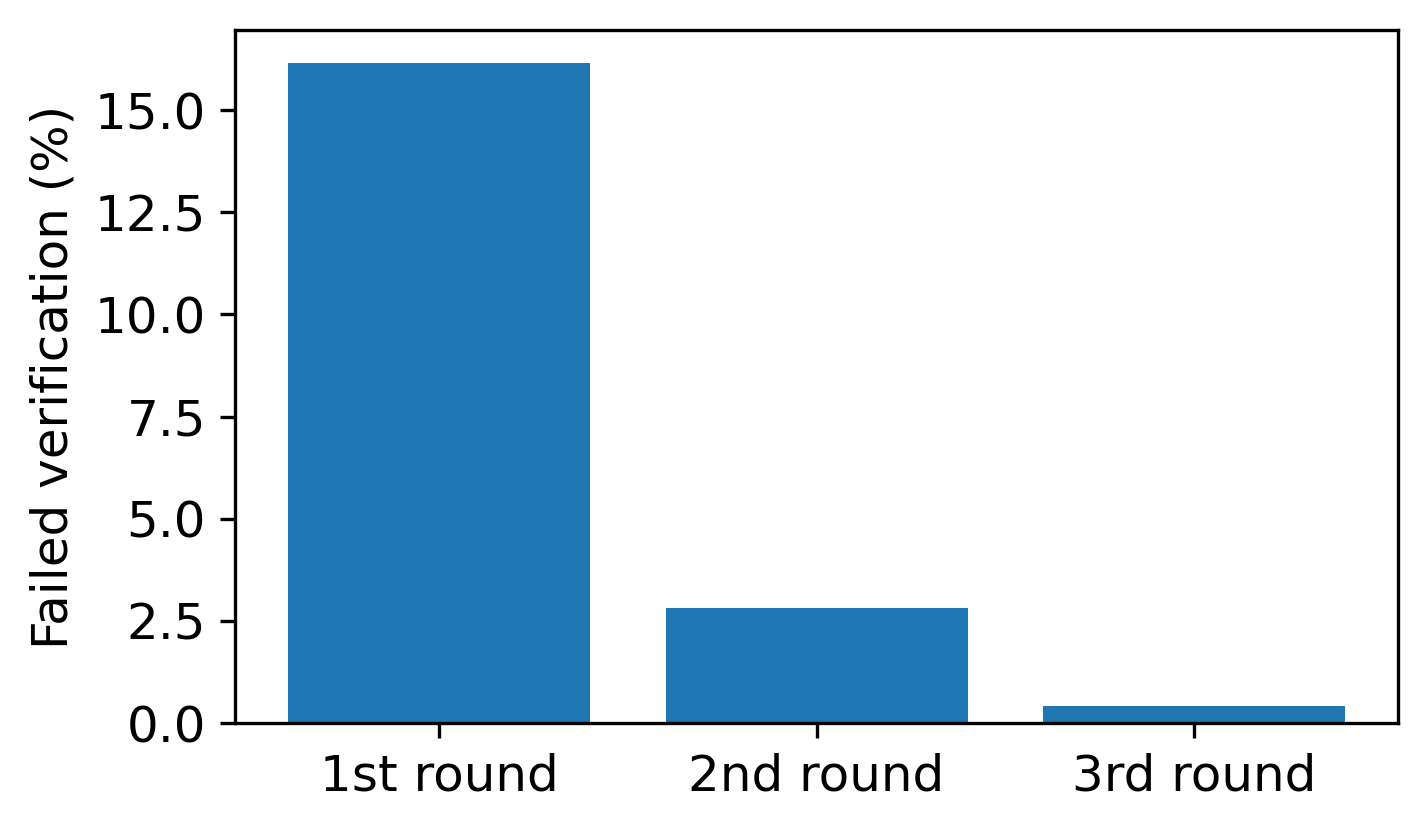}
        \caption{Iterative expert correction and verification required up to 
        four rounds.}
        \label{fig:veri-conflicts}
    \end{subfigure}\hspace*{\fill}
    \begin{subfigure}{0.32\textwidth}
        \includegraphics[width=\linewidth]{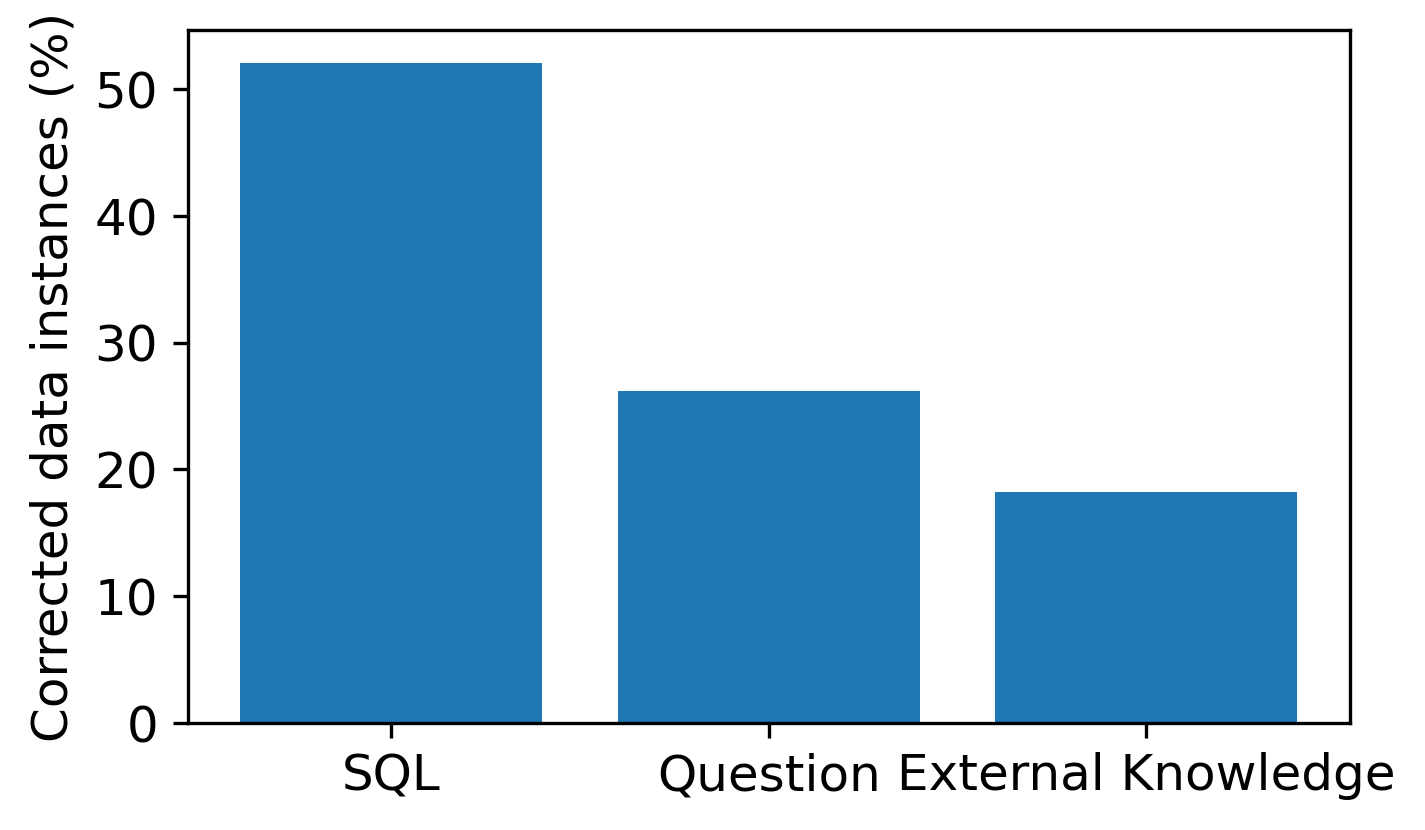}
        \caption{Pervasive annotation errors. Over 52\% of original gold SQL 
        queries are incorrect.}
        \label{fig:error-types}
    \end{subfigure}
    \caption{Quantitative analysis of the \dn expert curation process. Relying 
    solely on automated LLMs for data correction is insufficient due to poor 
    recall (Fig.~\ref{fig:llm-reviewer-metrics}). Consequently, \dn uses an 
    expert verification pipeline that requires up to four iterative rounds to fully 
    resolve errors (Fig.~\ref{fig:veri-conflicts}). We identified and corrected 
    errors across 52.1\% of SQL queries, 26.2\% of natural language questions, 
    and 18.2\% of external knowledge contexts (Fig.~\ref{fig:error-types}).}
    \label{fig:curation-stats}
\end{figure*}

\subsubsection{Schema Errors}

We corrected the schema information for 9 (1.7\%) tables to maintain consistency
with the example databases. Our corrections fall into the following two types.
\begin{enumerate}[leftmargin=*]
\item Removing references to tables or columns that do not exist in the provided 
example databases.
\item Correcting value descriptions that contradict the actual data. In the
example below, applying the original description would imply that a person
can be both a professor and a student at the same time:
\begin{tcolorbox}[
        colback=gray!5!white, 
        colframe=gray!20!white, 
        arc=4pt, 
        boxrule=1pt, 
        left=3pt, right=3pt, top=3pt, bottom=3pt,
        breakable
    ]
    \footnotesize
    \textit{A sample of the table person}:
\begin{verbatim}
p_id|professor|student|...
   3|        0|      1|...
   4|        0|      1|...
\end{verbatim}

    \vspace{0.5em}
    \textit{Original Value Explanation}:
\begin{verbatim}
column, value_description
professor, [...] 0: professor. 1: student.
student, [...] 0: professor. 1: student.
\end{verbatim}

    \vspace{0.5em}
    \textit{Corrected Value Explanation}: 
    \begin{verbatim}
column, value_description
professor, [...] 1: professor. 0: student.
student, [...] 0: professor. 1: student.
\end{verbatim}
    \end{tcolorbox}
\end{enumerate}

\subsubsection{Grading-Method Errors}

We find that the standard set-based grading method used in BIRD is insufficient
for two classes of questions. In addition to set-based grading, we annotated
instances with the following alternative grading methods:
\begin{enumerate}[leftmargin=*]
    \item \underline{Subset-based.} For any-k questions, such as  ``Please list 
    the titles of any two papers that Jundu has written'', we modified the gold
    query to return the full set without size limitation. Then, we check 
    whether the result set of a generated SQL query is a subset of that of the 
    gold SQL query with a correct size.
   
    \item \underline{List-based.} For questions requiring listing rows in a 
    specific order, such as ``List the product ID of the top five products, by 
    descending order, the number of quantities ordered'', we check whether the 
    result table of a generated SQL query has the same values as that of the 
    gold query at the same table cell position.
\end{enumerate}

\subsection{Construction and Dataset Statistics}
\label{subsec:con-stats}
We then describe the summary statistics for our data curation process and our 
curated dataset \dn.

\minihead{Construction process.}
In Figure \ref{fig:curation-stats}, we highlight the effectiveness of the LLM 
reviewer, the necessity of the involvement of SQL experts, and the importance of 
multi-round correction and verification. As shown in Figure 
\ref{fig:llm-reviewer-metrics}, the LLM reviewer achieves a high precision 
of 90.6\% in terms of detecting annotation errors, indicating that the errors
identified by the LLM reviewer are highly reliable. However, the LLM reviewer 
achieves a recall of only 24.5\%, missing a significant number of errors, 
which emphasizes the importance of expert involvement in data correction.

As shown in Figure \ref{fig:veri-conflicts}, we find that 16.2\% of the 
first-round proposed corrections fail the subsequent verification, 
indicating the necessity of multi-round correction and verification. In this 
process, we encountered five cases (0.2\%) where conflicts occurred between 
correction and verification, which were all resolved at the conflict-resolution stage.
Eventually, all the proposed corrections passed verification after up to four 
rounds.

\minihead{Dataset characteristics.}
In Figure \ref{fig:error-types} and Table \ref{tab:data-stats}, we summarize 
the statistics of our corrections and a comparison between the original BIRD 
Train and our \dn. As shown in Figure \ref{fig:error-types}, our process 
corrected gold SQL queries in 52.1\% of instances, natural language questions in 
26.2\%, and external knowledge in 18.2\%. Additionally, we discarded 1.5\% of 
instances whose questions are unanswerable given the database schema. In total, 
61.1\% of the evaluated instances contained at least one type of annotation error.

In Table \ref{tab:data-stats}, we show the average number of tables, joins,
functions, aggregations, set operations, subqueries, common table expressions 
(CTEs), and window functions per query. We find that our corrected queries have 
higher average counts in all categories, indicating that the corrected gold 
queries exhibit greater structural complexity, reflecting a more rigorous 
adherence to domain constraints.

\section{\sn: Shaping RLVR for Effective \ts Learning}
\label{sec:method}

We propose \sn, a reward shaping method with two techniques tailored
to \ts problems. In this section, we identify two failure modes of standard
result-based RLVR rewards on \ts (\S\ref{subsec:diagnosis}) and then propose
new reward shaping technique that addresses them (\S\ref{subsec:shaping}).

\subsection{Diagnosing the Failure Modes of Result-Based RLVR Rewards}
\label{subsec:diagnosis}

We discuss the problem setting and the RLVR formulation our method builds on and
then diagnose where standard rewards fail.

\minihead{BIRD-style \ts problems.}
We target the \ts setting introduced by BIRD~\cite{li2023can}. A 
\emph{BIRD-style \ts problem} provides (i) a natural language question, (ii) a 
relational database, given by its schema and contents, and (iii) a set of 
\emph{external knowledge} entries. The goal is to produce a SQL query that, 
when executed against the database, returns the answer to the question. The 
external knowledge, termed \emph{evidence} in BIRD, consists of short 
natural-language annotations that are ``required to map the natural language 
instructions into counterpart database values''~\cite{li2023can}. Because a
question is often under-specified without this evidence, a \ts model is expected
to read the external knowledge and reflect it in the generated query.

\minihead{RLVR background.}
We train models for this setting with RLVR, the dominant method for 
post-training LLMs \cite{guo2025deepseek,openaio3}. Group relative policy 
optimization (GRPO) \cite{shao2024deepseekmath} is one of the most widely used
algorithmic instantiations of RLVR. In each training iteration, for a 
question-answer pair ($q, a$), GRPO samples a group of rollouts 
$\{\tau_1, \ldots, \tau_G\}$ from the model $\pi_{\theta}$ and estimates 
the advantage of output $\tau_i$ as follows:
\begin{equation}
    \hat A_{i} = \frac{R(\tau_i, a) - \mathrm{mean}\left(\left\{R(\tau_j, a)\right\}_{j=1}^G\right)}{\mathrm{std}\left(\left\{R(\tau_j, a)\right\}_{j=1}^G\right)} 
    \label{eq:advantage}
\end{equation}
where $R$ is a task-specific reward function. GRPO then maximizes a
clipped surrogate objective over these advantages. The choice of $R$ thus
fully determines what the algorithm optimizes for.

Prior RLVR work for \ts \cite{yao2025arctic,databricks-rlvr,ma2025sql} uses a 
binary result-based reward, setting $R(\tau, a) = 1$ if executing the 
generated SQL against the example database produces the same result as the gold 
SQL query, and $0$ otherwise. While this formulation is cheap to compute during 
training, we show that such reward signals corrupt \ts learning in two 
systematic ways that limit the achievable accuracy of RLVR.

\begin{figure}
    \centering
    \begin{minipage}{0.46\linewidth}
        \includegraphics[width=\linewidth]{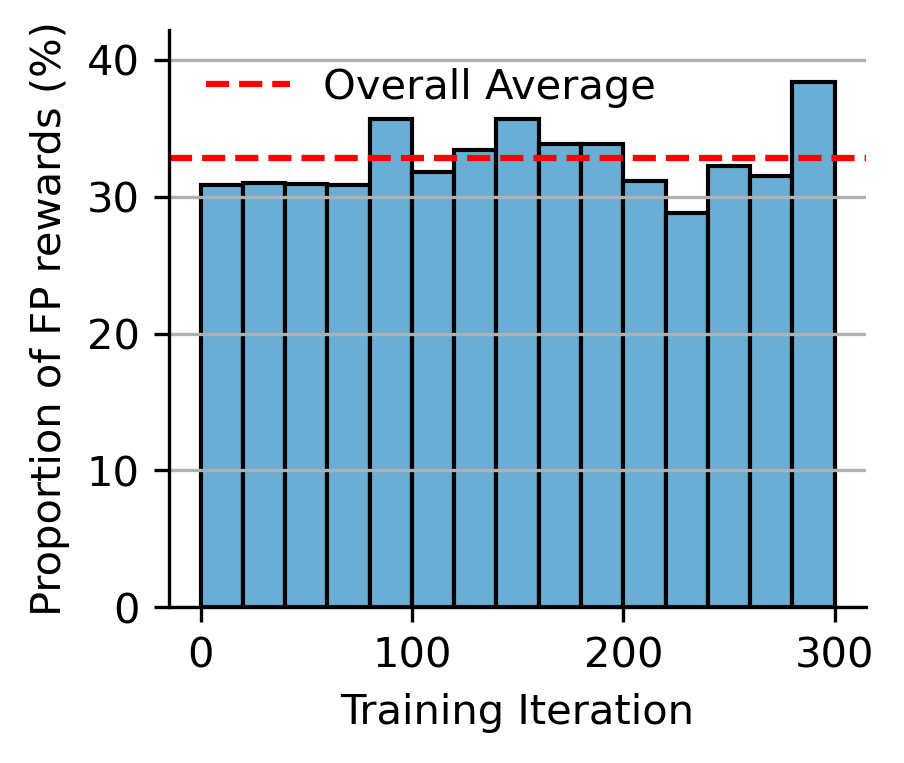}
        \caption{False positive rewards persist across the entire training
        process of Kimi-K2.6.}
        \label{fig:fp-rewards}
    \end{minipage}\hfill
    \begin{minipage}{0.48\linewidth}
        \includegraphics[width=\linewidth]{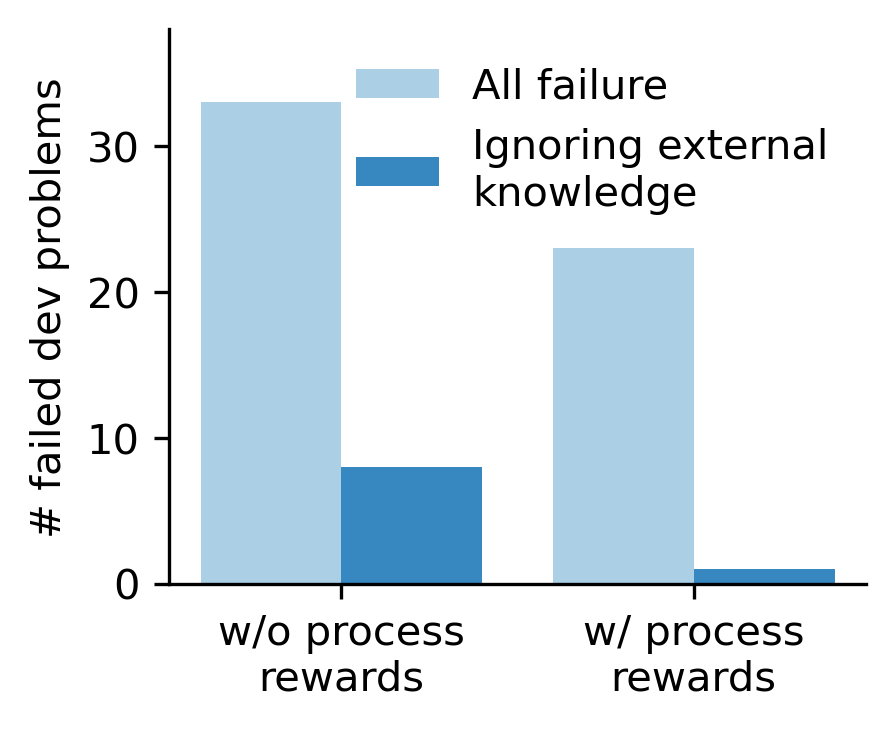}
        \caption{Process rewards reduce the failure cases caused by ignoring 
        external knowledge entries.}
        \label{fig:evi-sup}
    \end{minipage}  
\end{figure}

\minihead{Failure mode 1: false positive rewards from result coincidence.}
It is known that execution equivalence on a specific database does not guarantee 
semantic equivalence between two SQL queries \cite{chen2025parseval,he2024verieql}.
A query with an incorrect join, a missing predicate, or an accidentally 
compensating constant can produce the gold result on an example database at
training time. If such queries receive positive rewards, the model is pushed 
toward systematic logical errors that happen to be consistent with the example
database.

We measured the proportion of false positive rewards across the training process
on Kimi-K2.6 by applying VeriEQL \cite{he2024verieql} which verifies semantic 
SQL equivalence under bounded inputs (Appendix~\ref{appendix:verieql}). A reward 
is false positive if result-based grading accepts the query, while VeriEQL 
refutes its semantic equivalence to
the gold. During our pilot experiments (300 training steps), VeriEQL
identifies 28.8--38.4\% of positive rewards as false positives, with an average
rate of 32.8\% (Figure \ref{fig:fp-rewards}). Furthermore, this rate does not
decrease as training progresses and the model improves. The example below 
illustrates a typical case. The gold query averages over order totals, while the 
generated query averages over individual line-item prices. The two queries compute 
different quantities in general, but coincide on the example database because 
the customer happens to have orders containing a single line item each.

\begin{tcolorbox}[
        colback=gray!5!white, 
        colframe=gray!20!white, 
        arc=4pt, 
        boxrule=1pt, 
        left=3pt, right=3pt, top=3pt, bottom=3pt,
        breakable
    ]
    \footnotesize
    \textit{Question}: How much money on average does Lucas Wyldbore spend on book orders?

\lstset{escapeinside={<@}{@>}}

    \vspace{0.5em}
    \textit{Gold SQL Query}: 
\begin{lstlisting}
SELECT AVG(order_total)
FROM (
    SELECT T1.order_id, SUM(T1.price) AS order_total
    FROM order_line AS T1 JOIN cust_order As T2 ... 
    GROUP BY T1.order_id
);
\end{lstlisting}

    \vspace{0.5em}
    \textit{Incorrect Generated SQL Query getting positive rewards}: 
\begin{lstlisting}
SELECT AVG(price)
FROM order_line AS T1 JOIN cust_order As T2 ... 
-- same as the gold except for no GROUP BY
\end{lstlisting}
    \end{tcolorbox}

\minihead{Failure mode 2: outcome rewards do not enforce external-knowledge use.}
BIRD-style problems provide an evidence field as domain-specific external
knowledge. Standard outcome rewards based only on final execution results
provide no signal about how the model arrived at its answer. Therefore, a model
trained with outcome-only rewards may learn to ignore the external knowledge,
reverting to its pre-training priors over domain semantics.

To assess whether this failure mode appears empirically, we conducted a pilot 
manual analysis of the failures in a validation set for a model (Kimi-K2.6) trained with 
standard outcome-based RLVR. We treat a failed instance as ``ignoring external 
knowledge'' when the generated query is consistent with a plausible alternative 
interpretation of the question that the external knowledge is meant to rule out.
Among 33 failures, 8 instances (24.2\%) meet this criterion (Figure \ref{fig:evi-sup}, left bars). The example below illustrates a representative case.
The external knowledge specifies ``sodium-free refers to \texttt{sodium=0},'' 
while the model chooses the threshold of \texttt{sodium < 5}.

\begin{tcolorbox}[
        colback=gray!5!white, 
        colframe=gray!20!white, 
        arc=4pt, 
        boxrule=1pt, 
        left=3pt, right=3pt, top=3pt, bottom=3pt,
        breakable
    ]
    \footnotesize
    \textit{Question}: Among the recipes from The California Tree Fruit Agreement, calculate the percentage of sodium-free recipes.

    \vspace{0.5em}
    \textit{External knowledge}: The California Tree Fruit Agreement is a source; calculation \texttt{= MULTIPLY(DIVIDE(COUNT(sodium = 0 THEN recipe\_id), COUNT(recipe\_id)), 100)}
\lstset{escapeinside={<@}{@>}}

    \vspace{0.5em}
    \textit{Gold SQL Query}: 
\begin{lstlisting}
SELECT CAST(SUM(CASE WHEN T2.sodium = 0 THEN 1 ELSE 0 END) AS REAL) * 100 / COUNT(*) FROM ...
\end{lstlisting}

    \vspace{0.5em}
    \textit{Generated SQL Query that fails due to ignoring external knowledge}: 
\begin{lstlisting}
SELECT CAST(SUM(CASE WHEN n.sodium < 5 THEN 1 ELSE 0 END) AS REAL) * 100 / COUNT(*)FROM ...
\end{lstlisting}
    \end{tcolorbox}

Unfortunately, outcome rewards are structurally incapable of fixing the failures
due to ignoring rules in the external knowledge. By construction, outcome
rewards depend only on the final execution result, not on the reasoning that
produced it. An attempt that ignores
the external knowledge but coincidentally returns the correct result and an 
attempt that uses the external knowledge correctly are indistinguishable to an 
outcome reward. This observation motivates the process reward component
of \sn, which we will describe in the next section.

\subsection{Reward Shaping in \sn}
\label{subsec:shaping}

In \sn, we develop a composite reward that complements the result-based reward and
targets the two failure modes diagnosed above. For a generated rollout $\tau$ with
gold answer $a$, the \sn reward is
\begin{equation}
\begin{split}
& R_{\sn}(\tau, a) \\ &\;=\; \underbrace{\rho(\tau, a, \beta)}_{\substack{\text{shaped} \\ \text{outcome reward}}} \;-\; \underbrace{\sum_{k \in \mathcal{K}} \lambda_k \cdot \mathbbm{1}[\text{violation}_k(\tau)]}_{\substack{\text{external knowledge} \\ \text{process penalties}}}
\end{split}
\label{eq:revisql-reward}
\end{equation}
where $\rho$ is a verification-aware outcome reward that mitigates false
positives with penalty $\beta$, and $\mathcal{K}$ is a set of rules for external-knowledge use whose penalties $\lambda_k$ are deducted when violated. We describe 
each component as follows.

\minihead{Verification-aware outcome reward.}
To address Failure Mode 1, we integrate VeriEQL \cite{he2024verieql} with result-based 
grading as a second-stage check. Let $g(\tau, a) \in \{\text{True}, \text{False}\}$ denote the 
result-based verdict and 
$v(\tau, a) \in \{\text{True}, \text{False}, \bot\}$ denote VeriEQL's verdict, 
where $\bot$ indicates a VeriEQL timeout or a SQL query unsupported by VeriEQL. The 
shaped outcome reward is
\begin{equation}
\rho(\tau, a, \beta) =
\begin{cases}
0, & g = \text{False}, \\
1, & g = \text{True} \,\wedge\, v \in \{\text{True}, \bot\}, \\
1-\beta, & g = \text{True} \,\wedge\, v = \text{False}.
\end{cases}
\label{eq:outcome-reward}
\end{equation}

We downweight rather than zero out rewards that VeriEQL refutes because VeriEQL
itself can fail due to timeouts or unsupported SQL operators (\eg, window functions) 
\cite{he2024verieql}. Therefore, soft downweighting preserves the equivalence 
signal where VeriEQL is reliable while bounding the damage where it is not. 

\minihead{Process supervision for external knowledge use.}
To address Failure Mode 2, we require the model to explicitly translate and 
verify each external knowledge entry as part of its reasoning trace. We curate
a system prompt that requires two enumerated blocks indexed by the position of an external knowledge entry:
\begin{enumerate}[leftmargin=*]
    \item a ``requirement'' block that translates each external knowledge entry into a concrete 
    query constraint.
    \item a ``verification'' block right before the final answer that audits 
    whether the generated SQL query satisfies each external knowledge entry.
\end{enumerate}
We deduct penalties $\lambda_k$ when the model generates a solution without
complying with these requirements. We
conducted a pilot evaluation on the dev set to validate our reward shaping
intervention. With process rewards added, only 1 of 23 failed instances (4.3\%) 
is attributable to ignoring external knowledge. We defer the full ablation results on benchmarks to 
Section \ref{sec:eval-headline}.

\subsection{Training Algorithm and Implementation}
\label{subsec:impl}

Given the reward specification, we fine-tune models using the state-of-the-art (SOTA) RLVR 
settings \cite{liu2025skyrlsql} and loss function \cite{chen2025minimax,yu2025dapo}.

\minihead{Prompt.}
Our input prompt comprises two components: a universal system prompt shared by all
\ts problems (Appendix \ref{sec:app-prompts}), and the database schema of each
problem. We render the schema in data definition language and, following prior
work \cite{li2025omnisql,liu2025skyrlsql}, annotate each column with the column
description provided by BIRD. When a column description is unavailable, we instead
include three randomly sampled values from that column.

\minihead{Rollout.}
During training, the model performs multiple rollouts to generate 
multiple SQL queries. Within a rollout during training and inference, the model is allowed to issue intermediate SQL
queries using the provided SQL tool to explore data, test queries, and refine 
answers. We configure a maximum number of turns in which the model can issue intermediate queries
during a rollout. To aid the model's planning, we explicitly append a 
turn counter to the environment's response after each intermediate execution. We 
defer a detailed specification of a model rollout to Appendix \ref{sec:app-prompts}.

\minihead{Loss function.}
We use CISPO \cite{chen2025minimax}, a SOTA loss function of RLVR 
algorithms that stabilizes training and improves sample efficiency. Compared to GRPO,
CISPO introduces a constraint that clips gradients asymmetrically to prevent
token-level instability and mitigate the extreme variance caused by off-policy distribution shifts 
\cite{chen2025minimax}.

\minihead{Hyperparameters.}
We use a consistent set of hyperparameters as follows. We split \dn with an 
85:15 ratio for training and validation. For training, we use a group size of 16, 
a LoRA rank of 32, a batch size of 64, and a learning rate of $5\times 10^{-5}$. 
During rollouts, we allow up to five turns of interaction with the database for the model to issue intermediate queries,
given that our pilot experiments show that the average number of turns fluctuates 
between 3 and 4. For each turn, we follow prior work to allow the model to output at most 3,072 tokens \cite{liu2025skyrlsql}. For the composite reward computation, we set $\beta=0.2$
for verification-aware rewards and $\lambda_k=0.1$ for process supervision. We 
determine the training duration based on the convergence of the validation 
accuracy and select the checkpoint with the highest validation accuracy for the 
evaluation on benchmarks.

\section{Evaluation}
\label{sec:eval}

We assess the contributions of \dn and \sn across four benchmarks, addressing 
three questions. First, does \sn achieve human-level accuracy on BIRD, and how 
does it compare to the strongest prior open-source pipelines and models
(Section \ref{sec:eval-headline})? Second, how do verified data and reward shaping
each contribute to this result (Section \ref{sec:eval-headline})? Third, does
the benefit of \dn generalize beyond BIRD-style problems (Section 
\ref{sec:eval-generalization})? We first describe the experimental setup 
(Section \ref{sec:eval-settings}).

\subsection{Experimental Setup}
\label{sec:eval-settings}

We introduce our experimental settings, including benchmarks, baselines, 
metrics, and implementation details.

\minihead{Benchmarks.}
We evaluate on four benchmarks spanning two classes of \ts problems:
\begin{enumerate}[leftmargin=*]
\item \textit{Arcwise-Plat-Full}: As the original BIRD benchmark contains 
substantial annotation errors 
\cite{pourreza2023evaluating,jin2026pervasive,bird-error-panel,
wretblad2024understanding}, the Arcwise company introduced a manually corrected version 
of the BIRD Mini-Dev set (498 problems) \cite{Arcwise-minidev}. In addition, 
prior work \cite{jin2026pervasive} leveraged SQL experts and LLMs to fix extra 
errors in the Arcwise-corrected version and introduced Arcwise-Plat-Full.
\item \textit{Arcwise-Plat-SQL}: In addition to Arcwise-Plat-Full, prior work \cite{jin2026pervasive}
also introduced Arcwise-Plat-SQL, a variant that only corrects errors in the 
gold SQL while intentionally keeping the noise in natural language questions and 
external knowledge unchanged. Arcwise-Plat-SQL reflects the noise that often 
appears in real-world deployment \cite{jin2026pervasive}.
\item \textit{Spider2-SQLite}: As a successor of the widely used Spider1 
benchmark \cite{yu2018spider}, Spider2 consists of 547 challenging \ts problems 
\cite{lei2024spider}. Following prior work \cite{li2025omnisql}, we use 135 
problems of Spider2 that are based on SQLite. Gold SQL queries in Spider2 are more 
complex than those in BIRD, containing 5.2$\times$ more tokens on average 
\cite{lei2024spider}.
\item \textit{Spider2-Snow}: As a variant of Spider2, Spider2-Snow 
rewrites all 547 gold queries using the Snowflake dialect and requires the model to write Snowflake queries for evaluation \cite{lei2024spider}. 
\end{enumerate}

\minihead{Baselines.}
We compare against the five strongest open-source pipelines on the BIRD 
leaderboard as of May 1, 2026, four open-weight \ts models, and one proprietary reasoning model. 
We configured each pipeline with its leaderboard-optimal setting. If the 
leaderboard-optimal setting is not published, we use the default settings 
described in their technical report. For single-model baselines, we
use the prompt described in their technical report. If they do not specify a
prompt template, we apply the same prompts as \sn
(Appendix~\ref{subsec:app-sn-prompt}). We summarize the settings of baselines as follows.

\begin{enumerate}[leftmargin=*]
\item \textit{Contextual-SQL} relies on a dual-model architecture. It uses 
Qwen2.5-Coder-32B-Instruct for candidate generation, and a fine-tuned 32B reward 
model (trained on the entire 9.5k-example BIRD Train set) to perform the final 
candidate selection \cite{agrawal2025text2sql}.
\item \textit{CSC-SQL} uses a fine-tuned model, XiYanSQL-QwenCoder-32B-2412, to 
generate candidate SQL queries and uses Qwen2.5-Coder-7B-Instruct to select a 
final query for each problem \cite{sheng2025csc}.
\item \textit{GenaSQL} requires a multi-API orchestration pipeline 
spanning GPT-4o, Gemini 1.5, and Cohere to handle schema linking, retrieval, 
and generation tasks. To standardize evaluation, we used the same complex 
workflow and applied OpenAI's text-embedding-3-small for retrieval and 
GPT-5.2 for all subsequent orchestration and generation steps 
\cite{donder2025cheaper}.
\item \textit{OpenSearch} proposes multi-stage retrieval and refinement, using
bge-large-en-v1.5 for few-shot example retrieval and GPT-4o for schema 
extraction, query generation, and refinement \cite{xie2025opensearch}. In our 
experiment, we used the same retrieval model and replaced GPT-4o with GPT-5.2.
\item \textit{SHARE} takes a baseline SQL query as input and refines it across 
an ensemble of three distinct, distilled LLMs dedicated to reasoning, schema 
linking, and iterative query fixing. In our evaluation setup, this complex 
external refinement pipeline is applied to baseline queries generated by GPT-5.2 
\cite{qu2025share}.
\item \textit{OmniSQL-32B} is a fine-tuned model based on Qwen2.5-Coder-32B-Instruct 
using a combination of synthetic data (2.5 million examples), 
the Spider Train set, and the BIRD Train set \cite{li2025omnisql}.
\item \textit{XiYanSQL-QwenCoder-32B} is a fine-tuned model based on 
Qwen2.5-Coder-32B over a closed-source dataset \cite{liu2025xiyan}.
\item \textit{Infly-RL-SQL-32B} is a fine-tuned model without publicly disclosed
information about its fine-tuning process and usage \cite{infly-sql}.\footnote{We 
contacted the authors to request instructions for using the model. However, we 
have not received a response yet.} We used the same inference setup as OmniSQL-32B.
\item \textit{Arctic-R1-7B} is an RLVR model based on
Qwen2.5-Coder-7B-Instruct over 28k examples from BIRD Train set,
Spider Train set, Spider Dev set, and synthetic data \cite{yao2025arctic}.
\item \textit{GPT-5.2} is a frontier reasoning LLM. We execute 
it using the same inference settings as our models.
\end{enumerate}

\minihead{Metrics.}
We use two metrics. First, following prior work \cite{li2023can,lei2024spider}, we
measure the accuracy of generated SQL queries using their execution accuracy with
set-based grading. Second, we measure the inference cost for each SQL query. For methods based on
proprietary models, we calculate the costs based on the official pricing. For 
methods involving open-source models, we estimate their token costs 
using the cheapest pricing of their corresponding base models provided by Together 
\cite{together-pricing}, Groq \cite{groq-pricing}, Sail \cite{sail-pricing},
or Fireworks \cite{fireworks-pricing}. We used pricing data from May 1,
2026. In addition, we compute confidence intervals and run statistical 
significance tests when comparing a method against the human-level proxy and 
defer the technical details of these statistical procedures to 
Appendix~\ref{appendix:stats}.

\minihead{Implementation details.}
For the training stage of \sn, we used Tinker \cite{tinker} as the 
infrastructure. Each intermediate query issued by the model is subject to a 
30-second timeout. We conducted our training experiments on a machine with an 
Intel 8568Y CPU, 512 GB memory, and ran open-source models on one H100.

\subsection{\sn Reaches Human-Level Accuracy on BIRD}
\label{sec:eval-headline}
\begin{figure*}[t!]
    \begin{subfigure}{0.24\linewidth}
        \includegraphics[width=\linewidth]{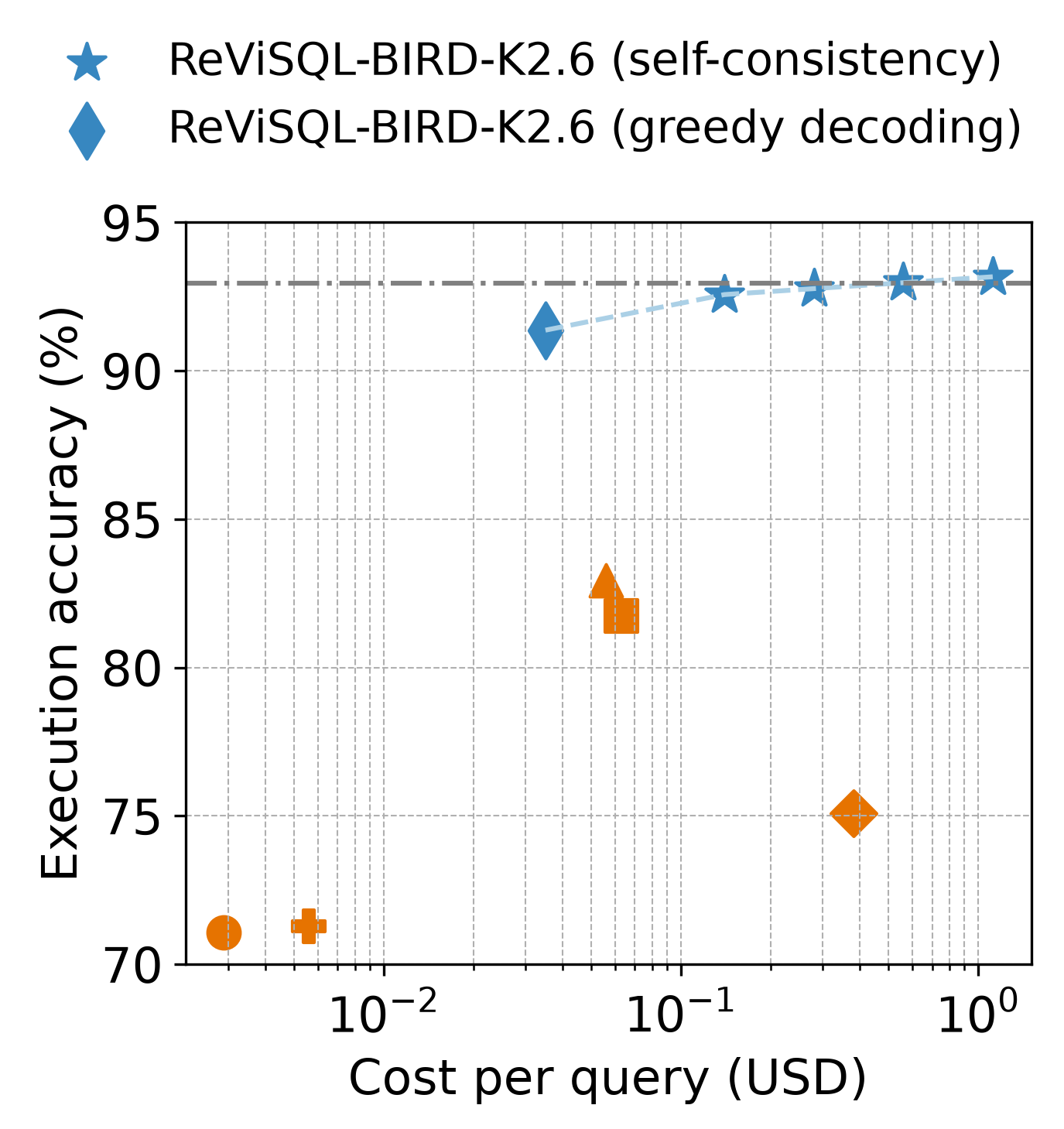}
        \caption{Arcwise-Plat-SQL.} 
        \label{fig:agents-sql}
    \end{subfigure}\hspace*{\fill}
    \begin{subfigure}{0.24\linewidth}
        \includegraphics[width=\linewidth]{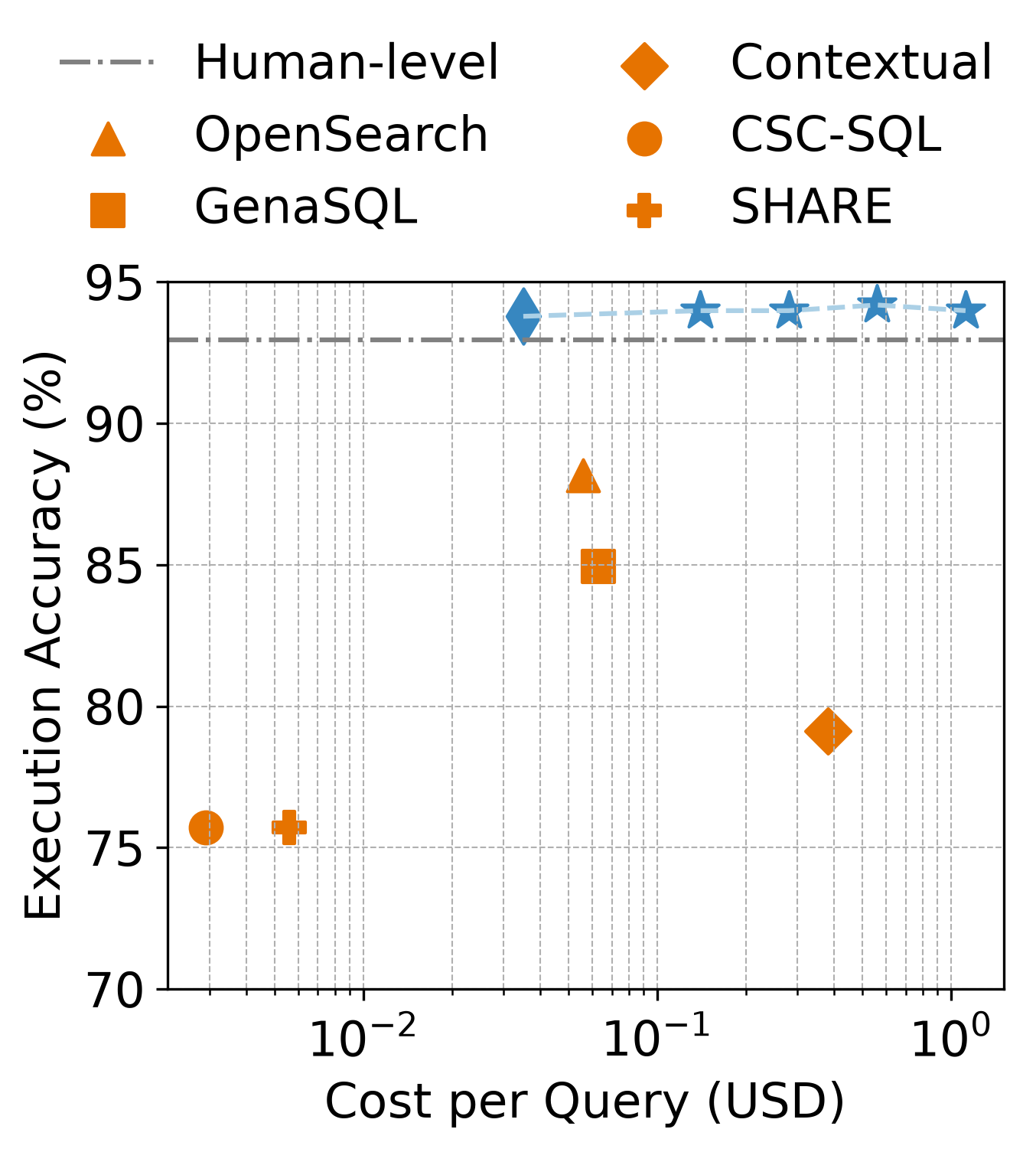}
        \caption{Arcwise-Plat-Full.}
        \label{fig:agents-full}
    \end{subfigure}\hspace*{\fill}
    \begin{subfigure}{0.24\linewidth}
        \includegraphics[width=\linewidth]{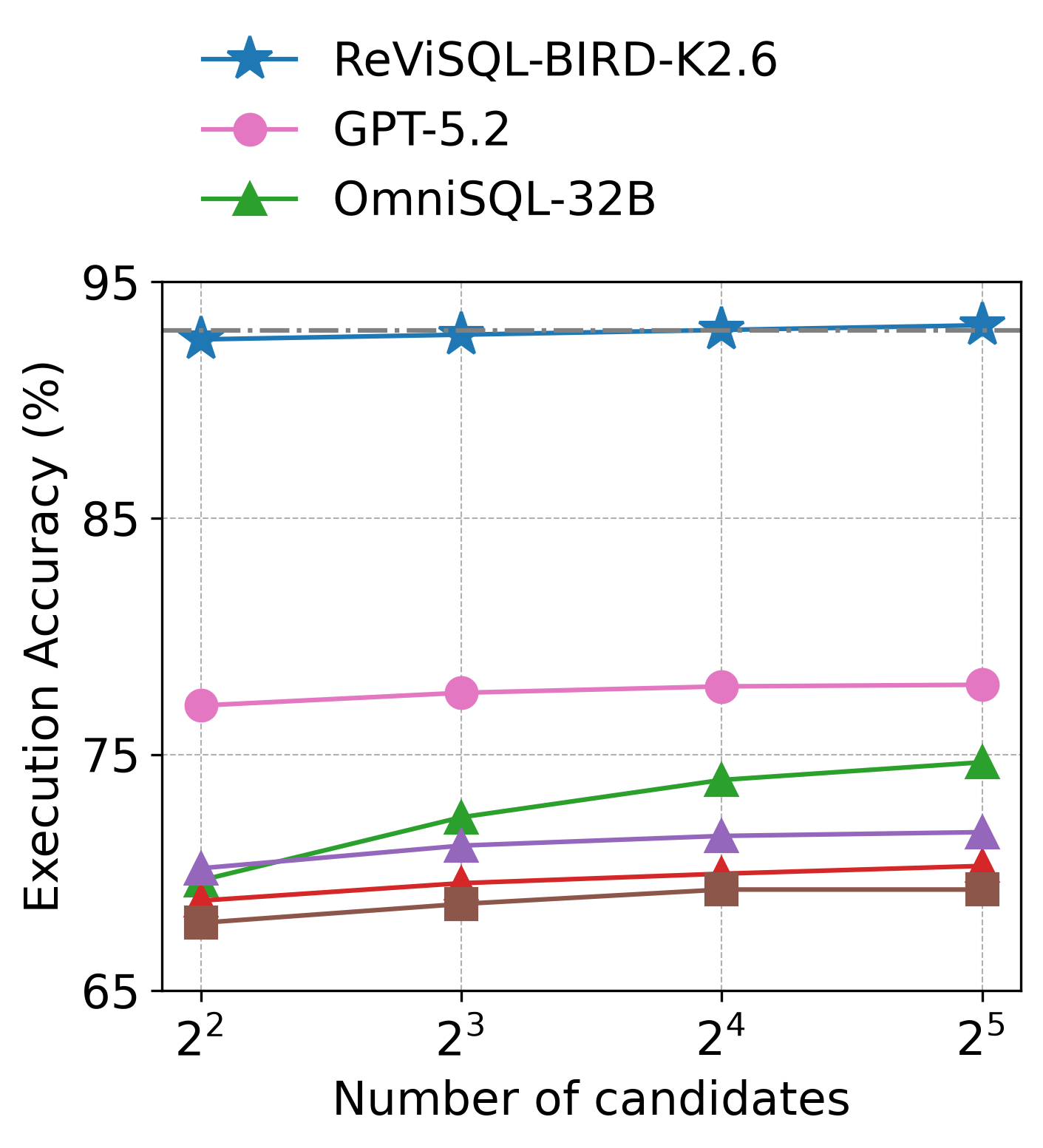}
        \caption{Arcwise-Plat-SQL.} 
        \label{fig:single-sql}
    \end{subfigure}\hspace*{\fill}
    \begin{subfigure}{0.24\linewidth}
        \includegraphics[width=\linewidth]{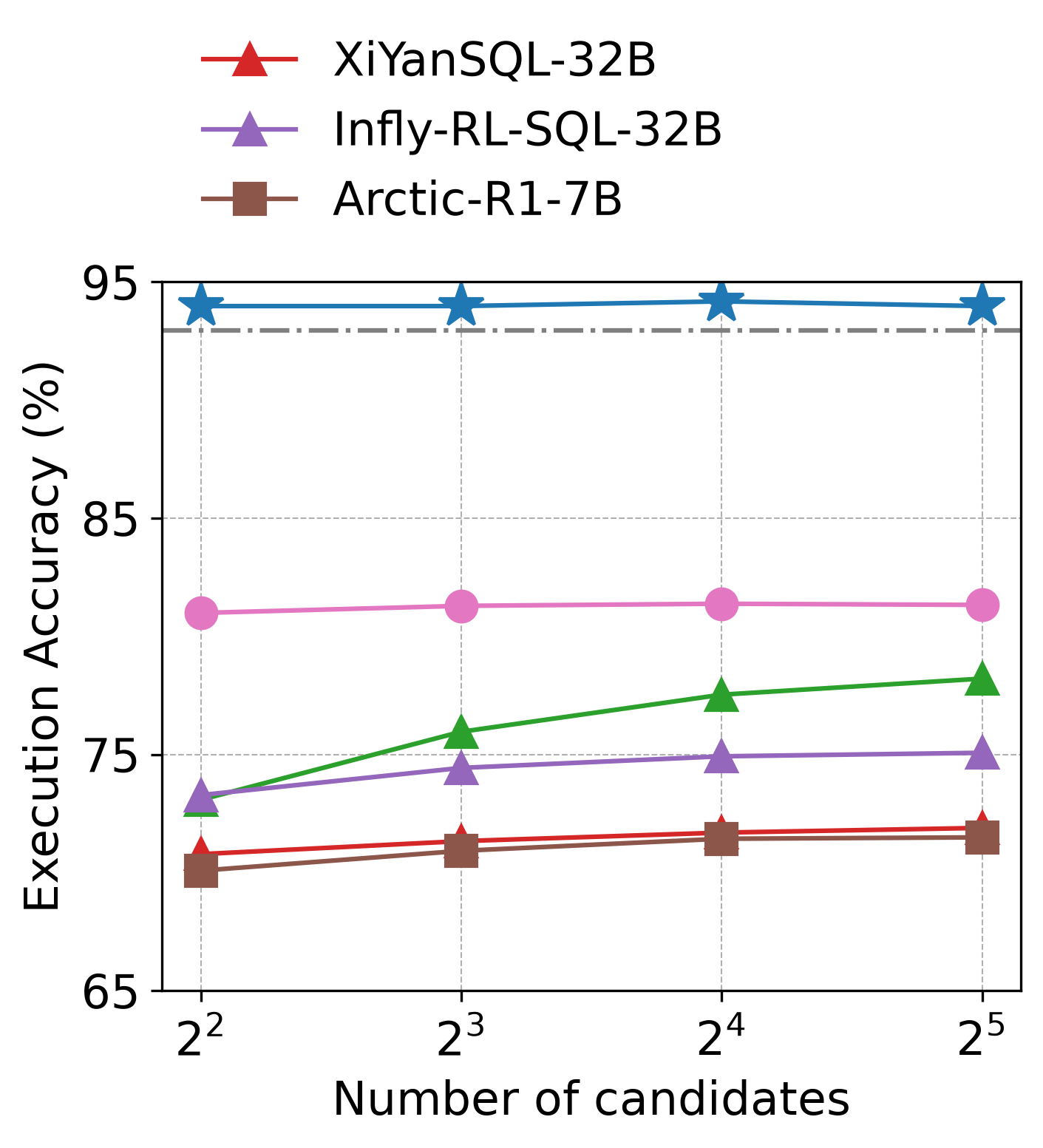}
        \caption{Arcwise-Plat-Full.}
        \label{fig:single-full}
    \end{subfigure}\hspace*{\fill}
    \caption{\sn-K2.6 achieves human-level accuracy on BIRD while 
    dominating prior methods. (a, b) Accuracy vs.\ cost on 
    Arcwise-Plat-SQL and Arcwise-Plat-Full, comparing \sn-K2.6 against 
    the top 5 open-source methods on BIRD Leaderboard. (c, d) Accuracy vs.\ 
    candidate count, comparing \sn-K2.6 against single-model baselines.}
    \label{fig:human-level}
\end{figure*}

We instantiated \sn on Kimi-K2.6 \cite{kimi26} and evaluated the resulting model,
\sn-K2.6, on Arcwise-Plat-SQL and Arcwise-Plat-Full under two decoding methods: 
(1) greedy decoding and (2) self-consistency over $n$ candidates. For 
self-consistency, we follow the standard procedure used in prior 
work~\cite{wang2022self,dong2023c3,gao2023text,xie2025opensearch,sheng2025csc}. 
We sample $n$ candidates with temperature 1 and top-$p$ 1, group them by 
execution results, and randomly return a query from the largest group, breaking 
ties uniformly at random. We repeat this process five times and report the 
average accuracy.

\minihead{\sn-K2.6 is the first method to achieve human-level
accuracy.} As shown in Figure \ref{fig:agents-full}, with greedy decoding, 
\sn-K2.6 achieves 93.8\% (95\% CI [91.3, 95.6]) on Arcwise-Plat-Full. As shown 
in Figure \ref{fig:agents-sql}, with self-consistency over 16 candidates, 
\sn-K2.6 achieves 93.0\% (95\% CI [90.4, 94.9]) on Arcwise-Plat-SQL. For both 
results, the human-level accuracy of 92.96\% \cite{bird-leaderboard} lies 
within the 95\% confidence interval. In contrast, every prior system we 
evaluate falls significantly below the human proxy (\eg, OpenSearch at 
82.9\%/88.2\%, $p<10^{-4}$). To our knowledge, \sn-K2.6 is the first \ts 
method to achieve human parity on expert-verified BIRD datasets.

\minihead{\sn-K2.6 dominates complex pipelines.} As shown in Figures 
\ref{fig:agents-sql} and \ref{fig:agents-full}, the prior best open-source 
pipeline, OpenSearch (based on GPT-5.2), achieves 82.9\% on 
Arcwise-Plat-SQL and 88.2\% on Arcwise-Plat-Full at a cost of \$0.056 per 
query. \sn-K2.6 with greedy decoding outperforms OpenSearch by 8.4\% on
Arcwise-Plat-SQL and 5.6\% on Arcwise-Plat-Full, \emph{while costing
37\% less per query}. Compared to other pipelines, \sn-K2.6 with greedy 
decoding achieves 9.6--20.3\% and 8.8--18.1\% higher accuracy, respectively. 
With self-consistency over up to 32 candidates, \sn-K2.6 outperforms all the 
complex pipelines by 9.6--22.1\% and 5.8--18.5\% on Arcwise-Plat-SQL and -Full, respectively.

\minihead{\sn-K2.6 outperforms open-weight and frontier models.}
In Figures~\ref{fig:single-sql} and \ref{fig:single-full}, we compare \sn-K2.6 
against the best open-weight \ts models and GPT-5.2, as the number of generated 
candidates varies from 4 to 32. Across all candidate budgets and both benchmarks, 
\sn-K2.6 achieves 12.7--24.7\% higher accuracy. Furthermore, \sn-K2.6 with 
4 candidates exceeds the performance of all baselines even when the latter are
given a budget of 32
candidates. This result indicates that test-time compute fails to compensate for weak
intrinsic model capabilities on \ts.

\begin{figure}
    \centering
    \includegraphics[width=\linewidth]{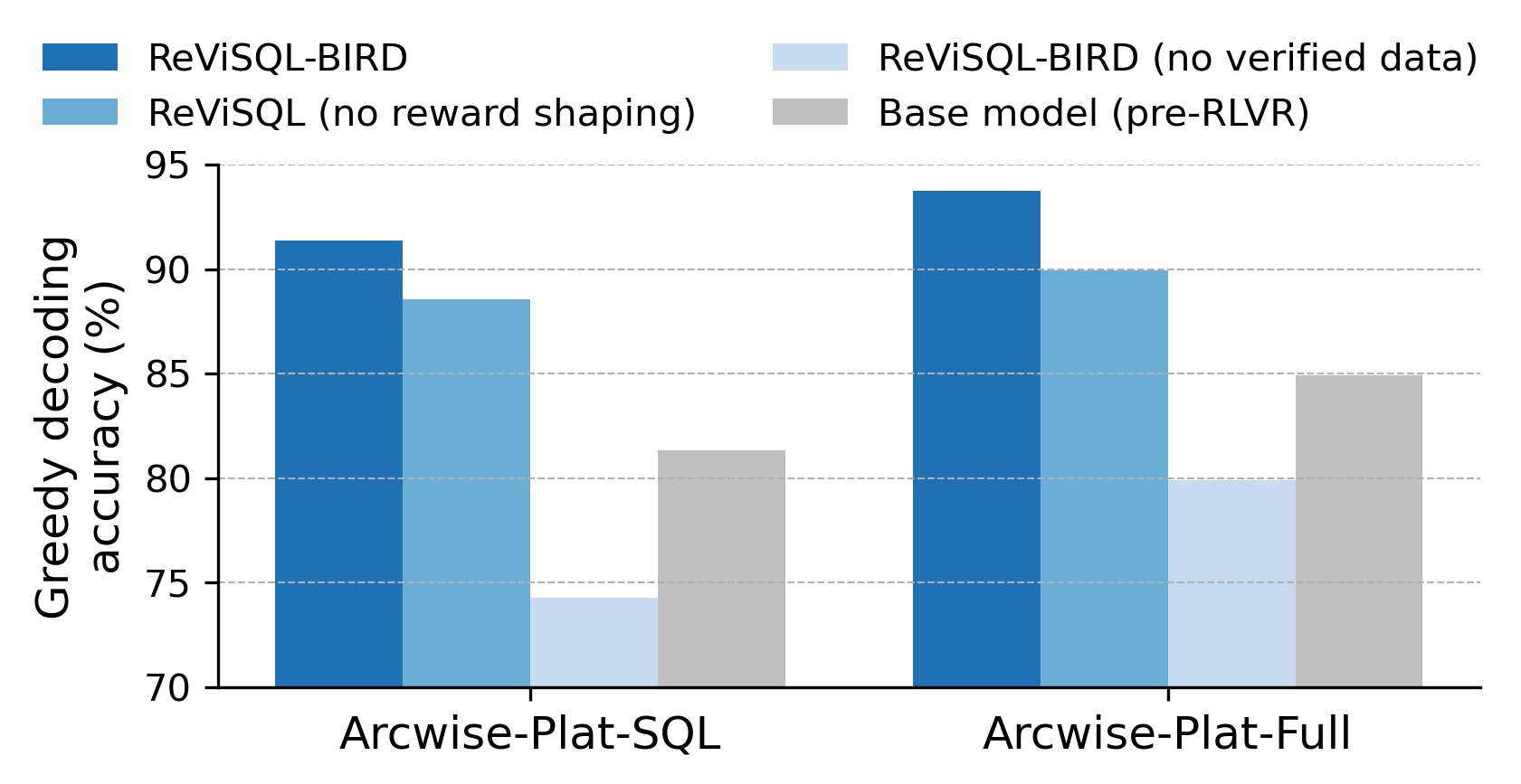}
    \caption{Ablation of \sn-K2.6 on Arcwise-Plat-SQL and -Full under greedy
    decoding. Training on \dn rather than 
    the original BIRD Train improves accuracy by 13.9--17.1\%. Reward shaping then further improves accuracy by 
    2.8--3.8\%, reducing error rates by 24.6--38.0\%.}
    \label{fig:ablation}
\end{figure}

\minihead{Verified data is a dominant factor while reward shaping bridges the
gap to human parity.}
To isolate the contribution of each ingredient, we ablate \sn-K2.6 under greedy
decoding on the Arcwise-Plat benchmarks (Figure~\ref{fig:ablation}), comparing
\sn against three configurations: (1) \rvbase (no reward shaping), (2) \sn on
BIRD Train, and (3) the base model with the same prompts as \sn. All 
configurations share the same hyperparameters and checkpoint-selection method 
discussed in Section \ref{subsec:impl}.

\begin{table*}[h]
    \centering
    \caption{\rvbase-235B (RLVR on BIRD-Platinum) outperforms the strongest 
    prior open-source pipelines across two BIRD-derived benchmarks 
    (Arcwise-Plat-SQL, Arcwise-Plat-Full) and the out-of-distribution Spider 
    2-SQLite benchmark, at lower per-query cost, demonstrating the generalizability of \dn.}
    \label{tab:spider2sqlite}
    \begin{threeparttable}
    \centering
    \begin{tabular}{l S[table-format=2.2] S[table-format=2.2] S[table-format=2.2] c}
        \toprule
        \textbf{Method} & \textbf{Arcwise-Plat-SQL (\%)} & \textbf{Arcwise-Plat-Full (\%)} 
        & \textbf{Spider 2-SQLite (\%)} & \textbf{Per-query cost (USD)} \\ 
        \midrule

        SHARE (GPT-5.2)             & 71.29 & 75.70 & 21.48 & $5.6 \times 10^{-3}$ \\ 

        CSC-SQL (XiYan-32B)     &  71.08 & 75.70 & 10.37 & $2.9 \times 10^{-3}$ \\ 
        
        
        Contextual-SQL (XiYan-32B)  & 75.10 & 79.12 & 11.11 &  $3.8 \times 10^{-1}$ \\
        
        GenaSQL (GPT-5.2)           & 81.73 & 84.94 & 34.81 & $6.3 \times 10^{-2}$ \\ 
        
        OpenSearch (GPT-5.2)    & 82.93 & 88.15 & 34.07\tnote{$*$} & $5.6 \times 10^{-2}$ \\ 
        
        
        
        

        
        \textbf{\rvbase-235B (greedy)} & \textbf{87.34} & \textbf{88.75} & \textbf{37.04} & \textbf{$7.7 \times 10^{-3}$} \\ 
        \bottomrule
    \end{tabular}
    \begin{tablenotes}
        \footnotesize
        \item[$*$] OpenSearch \cite{xie2025opensearch} relies on training data for dynamic few-shot prompting, which Spider 2 does not provide. We tested with substituting the BIRD and Spider 1 training sets, as well as disabling the module entirely. We report the performance with dynamic few-shot disabled, as it achieved the highest performance.
    \end{tablenotes}
    \end{threeparttable}
\end{table*}%

We find that verified data is the dominant driver. Although BIRD Train contains 
3.8$\times$ more instances than \dn, training on \dn improves the base model by 
7.2\% and 5.0\% on Arcwise-Plat-SQL and -Full, whereas training on BIRD Train 
lowers it by 7.0\% and 5.0\%. We find that this is because 
rewarding a model against erroneous gold queries teaches it to reproduce the 
annotation errors (\eg, the \texttt{DISTINCT} omission discussed in 
\S\ref{subsec:errors-sql}). 
Furthermore, reward shaping bridges the residual gap, improving the accuracy by
2.8\% and 3.8\%. While these absolute gains appear modest, they are decisive at
this accuracy level. Reward shaping removes 24.6\% and 38.0\% of the errors that
verified data alone leaves behind. It is what lifts \sn-K2.6 to human parity.

\minihead{The residual gap is dominated by question ambiguity.}
To characterize the residual gap, we manually inspected every \sn-K2.6 
self-consistency failure across both Arcwise-Plat benchmarks, classifying each 
as (1) question ambiguity, where the model produces a valid reading of a 
natural question that differs from the gold's, or (2) a genuine model failure. We 
find that 69\% of the cases are attributable to question ambiguity, and 31\% to model
defects. Specifically, Arcwise-Plat-SQL carries more ambiguity-driven failures 
than Arcwise-Plat-Full (24 vs.\ 20), which is consistent with its preserved 
noisy questions. Among genuine model failures, 60\% of the cases are due to omitting a
schema-induced filter (\eg, filtering \texttt{NULL} values). We provide the full
taxonomy, per-type analysis, and case studies in Appendix~\ref{appendix:residual}.

\subsection{\dn Generalizes Across Benchmarks}
\label{sec:eval-generalization}

We evaluate whether our curated training data, \dn, and our \rvbase recipe
(RLVR on \dn) generalize beyond BIRD. To isolate the data-quality contribution
from any algorithmic effect, we trained Qwen3-235B-A22B using \rvbase without the
reward shaping methods discussed in Section \ref{sec:method}. We then evaluated
\rvbase-235B on all four benchmarks.

\rvbase-235B is trained on \dn, where all SQL queries are in SQLite syntax. 
Therefore, evaluating it on Spider2-Snow, which only accepts Snowflake-dialect 
queries, requires two inference-time adaptations without re-training. First, we
translated the intermediate exploratory SQL queries issued by the model from
SQLite to Snowflake using SQLGlot~\cite{sqlglot}. When SQLGlot fails, we fall 
back to a minimal automatic conversion that quotes identifiers with double 
quotes. Second, because Spider2-Snow databases have schemas and questions whose
combined length exceeds our fine-tuned model's context window, we applied the
column- and table-filtering procedure from ReFoRCE~\cite{deng2025reforce} to
select schema elements relevant to each question. All other inference settings 
match our BIRD evaluation.

\begin{figure}
    \centering
    \includegraphics[width=\linewidth]{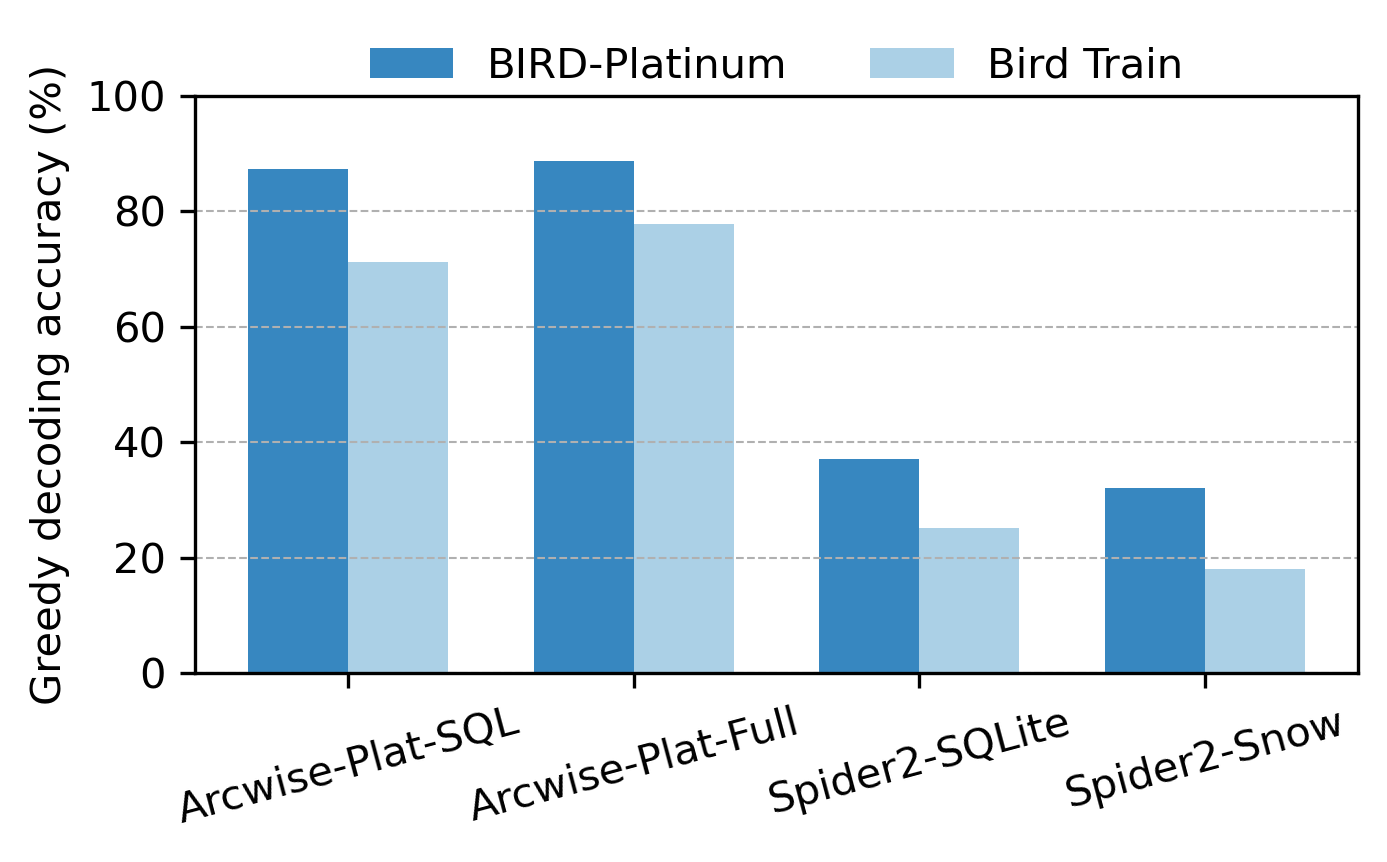}
    \caption{Training Qwen3-235B-A22B with RLVR on BIRD-Platinum (\rvbase)
    yields consistent accuracy improvements (11.0--16.1\%) over training on the original BIRD
    Train, across Arcwise-Plat and Spider2 benchmarks.}
    \label{fig:rvbase-generalization}
\end{figure}

\minihead{\rvbase improves accuracy on all benchmarks.} 
In Figure~\ref{fig:rvbase-generalization}, we compare \rvbase-235B trained on 
\dn against the same base model trained on the original BIRD Train, under greedy 
decoding. On the BIRD-derived Arcwise-Plat benchmarks, \rvbase-235B outperforms 
its BIRD Train counterpart by 16.1\% on Arcwise-Plat-SQL and 11.0\% on 
Arcwise-Plat-Full. Furthermore, the improvement persists on Spider2
benchmarks. \rvbase-235B achieves 11.8\% and 14.0\% higher accuracy on 
Spider2-SQLite and Spider2-Snow compared to training on BIRD Train. The 
improvements across benchmarks (different SQL query complexities and dialects) 
demonstrate that the data-quality effect is not specific to BIRD. RLVR with \dn 
produces \ts models that generalize more broadly.

\minihead{\rvbase-235B outperforms the open-source pipelines on
Spider2-SQLite.} In Table~\ref{tab:spider2sqlite}, we compare \rvbase-235B
against the top 5 open-source pipelines on the BIRD Leaderboard. On Spider2-SQLite, 
with greedy decoding, \rvbase-235B achieves 37.04\% accuracy at \$0.0077 per 
query, outperforming the previous best pipeline (GenaSQL) by 2.2\% at 8$\times$ 
lower cost, and outperforming OpenSearch by 3\% at 7$\times$ lower 
cost. \rvbase-235B also leads on Arcwise-Plat-SQL and Arcwise-Plat-Full, 
outperforming baselines by 0.6--16.3\%.

\minihead{BIRD-Platinum generalizes to Snowflake dialects.} On the more
challenging Spider2-Snow benchmark, which requires Snowflake-dialect SQL,
\rvbase-235B with self-consistency over 128 candidates achieves 55.58\% accuracy, outperforming agentic pipelines built on
substantially larger models, including ReFoRCE and AutoLink on DeepSeek
(671B) and Spider-Agent on Qwen3-Coder (480B), by 17.55\%, 0.74\%, and 24.5\% \cite{spider2-leaderboard}.
Since Spider2-Snow is held out from \rvbase's
BIRD-Platinum training data, this result provides evidence that \rvbase's 
improvements transfer to new SQL dialects.

\section{Conclusion}
We show that human-level \ts accuracy is achievable by fine-tuning
a single LLM with RLVR on clean data, without the complex pipelines that 
dominate top-ranking LLM-based methods. We introduce two advances in RLVR for 
\ts. First, we develop \dn, a 2.5k-instance verified dataset constructed through a 
hybrid LLM--human verification pipeline. We repair the annotation errors in 61\% 
of the sampled instances from BIRD Train, yielding 11--16\% accuracy improvements 
over training on BIRD Train under the same RLVR algorithm. Second, we introduce 
\sn, a specialized reward shaping method designed for the failure modes of 
RLVR in \ts tasks. We propose a verification-aware outcome reward
to mitigate the false positive rewards in standard result-based rewards, and 
leverage rule-based process rewards to prevent models from ignoring external knowledge. We trained \sn-K2.6, the first method to match human-level 
accuracy (92.96\%) on expert-verified variants of BIRD, outperforming top 
open-source methods by 10--22\%. 


\section*{Acknowledgments}
This research was supported in part by a sponsorship from Bridgewater AIA Labs. 
We are grateful to the CloudLab for providing computing resources for experiments 
\cite{Duplyakin+:ATC19}.

\clearpage

\bibliographystyle{ACM-Reference-Format}
\bibliography{main}

\clearpage

\appendix
\section{Case Study: The Two Expert-Verified Arcwise-Plat Benchmarks}
\label{appendix:arcwise-case-study}

We evaluate on two expert-verified variants of the BIRD Mini-Dev set released by
prior work~\cite{Arcwise-minidev,jin2026pervasive}: Arcwise-Plat-Full and
Arcwise-Plat-SQL. The two are built from the same instances but differ in which
sources of annotation error are corrected (Table~\ref{tab:arcwise}).
Arcwise-Plat-Full corrects all three sources: the natural language question,
the external knowledge, and the gold SQL query. Therefore, it measures \ts
accuracy under clean inputs. Arcwise-Plat-SQL corrects only the gold SQL query
and \emph{deliberately preserves} the original, often noisy, questions and
external knowledge. Hence, it measures robustness to the ambiguous inputs
that are typical of real-world \ts deployments~\cite{liu2025survey}. We walk 
through a representative instance to make the distinction concrete.

\begin{figure*}[t]
    \centering
    \includegraphics[width=\linewidth]{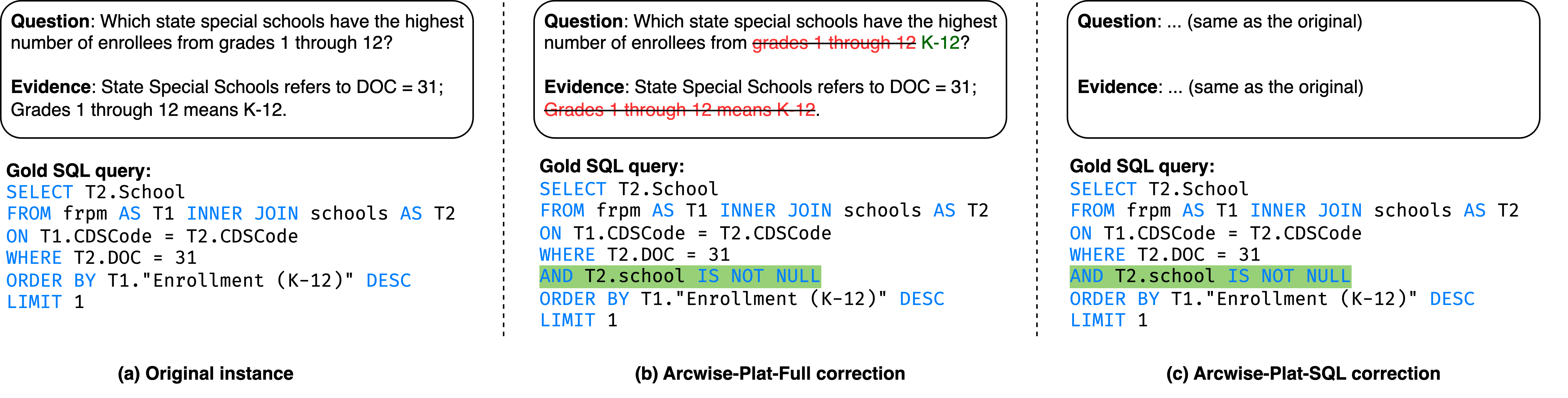}
    \caption{A representative instance of two expert-verified variants of BIRD 
    Mini-Dev. (a) The original instance contains an incorrect phrasing in the 
    question (``grades 1 through 12'' as opposed to ``K-12'') and a gold SQL 
    query that omits a \texttt{NOT NULL} filter, causing it to include data 
    about districts and schools. (b) Arcwise-Plat-Full corrects all three error 
    sources: rephrasing the question to ``K-12,'' removing the redundant 
    external-knowledge entry, and augmenting gold SQL. (c) Arcwise-Plat-SQL 
    corrects only the gold SQL, preserving the original noisy question and 
    external knowledge.}
    \label{fig:eval-correction-examples}
\end{figure*}

\minihead{A representative instance.}
In Figure~\ref{fig:eval-correction-examples}, we show one BIRD Mini-Dev instance 
under all three versions. The original instance 
(Figure~\ref{fig:eval-correction-examples}a) contains an error in each of the
three sources. The question asks about schools serving ``grades 1 through 12,''
an incorrect phrasing of the intended ``K-12'' criterion; the external knowledge
includes a redundant entry; and the gold SQL query omits a \texttt{NOT NULL}
filter, so its result wrongly includes rows about districts in addition to
schools.

\emph{Arcwise-Plat-Full} repairs all three
(Figure~\ref{fig:eval-correction-examples}b): it rephrases the question to
``K-12,'' removes the redundant external-knowledge entry, and augments the gold
SQL with the missing filter. A model evaluated here sees an unambiguous question
and is graded against a correct gold query, so the benchmark isolates the
model's SQL-reasoning ability.

\emph{Arcwise-Plat-SQL} repairs only the gold SQL
(Figure~\ref{fig:eval-correction-examples}c): the missing \texttt{NOT NULL}
filter is added, but the noisy ``grades 1 through 12'' question and the
redundant external knowledge are left untouched. A model must now recover the
intended ``K-12'' meaning from an imperfect question without the benefit of a
sanitized prompt. This is the harder, deployment-realistic condition: the input
is exactly as noisy as the original benchmark, yet the gold answer is now
correct, so a model is neither rewarded for matching an erroneous gold query nor
rescued by a cleaned question.

\minihead{Implication.}
Contrasting the two benchmarks isolates two distinct capabilities. A gap between
a method's Arcwise-Plat-Full and Arcwise-Plat-SQL accuracy reflects its
sensitivity to noisy natural-language inputs rather than its raw SQL competence.
This is consistent with the headline result in
Section~\ref{sec:eval-headline}: \sn-K2.6 reaches human-level accuracy on the
clean Arcwise-Plat-Full with greedy decoding, but recovers the same level on the
noisier Arcwise-Plat-SQL only with self-consistency over multiple candidates,
indicating that inference-time scaling chiefly compensates for question
ambiguity rather than for limited SQL reasoning.

\section{Detailed Analysis of Residual Failures}
\label{appendix:residual}

\lstset{
    language=SQL,
    basicstyle=\ttfamily\scriptsize,
    breaklines=true,
    breakatwhitespace=true,
    keepspaces=true,
    columns=fullflexible,
}

In this appendix, we report a case-by-case analysis of the failures that remain
after \sn-K2.6 reaches human-level accuracy (Section~\ref{sec:eval-headline}). 
We aim to answer the question: of the small residual gap, how much reflects 
genuine limitations of the model's SQL generation, and how much reflects 
inherent ambiguity in the natural-language questions? We treat the 
expert-verified Arcwise-Plat gold labels as correct throughout and attribute 
every disagreement either to question ambiguity or to a model defect.

\subsection{Scope and Method}

\minihead{Failure set.}
We collected, for every problem on which \sn-K2.6's self-consistency answer
disagreed with the gold, the single most frequent incorrect query the model
produced. This yields 64 cases spanning 43 distinct problems over 11 databases,
30 from Arcwise-Plat-Full and 34 from Arcwise-Plat-SQL (a problem may appear in
both benchmarks). For each case we examined the question, the external
knowledge, the gold query, the model's query, and the database schema.

\minihead{Two-axis taxonomy.}
We classify each case along two axes. The first is the source of the
disagreement: \emph{question ambiguity}, where the question admits more than one
reasonable reading and the model realizes a valid interpretation other than the
one the gold encodes; or a \emph{model defect}, where the model's query is
genuinely incorrect. The second axis is the \emph{mechanism}, refined into three
categories on each side (Table~\ref{tab:residual}). Two of the authors conduct 
the classification manually and reach agreement via discussion.

\subsection{Findings}

\begin{table}[t]
    \centering
    \caption{Classification of the 64 residual \sn-K2.6 failures (the most
    frequent incorrect self-consistency query on each problem where the model
    disagrees with the gold, across both Arcwise-Plat benchmarks). Most failures
    trace to ambiguity in the question rather than to defects in the model's SQL
    generation.}
    \label{tab:residual}
    \begin{tabular}{llr}
        \toprule
        Attribution & Mechanism & \# cases \\
        \midrule
        \multirow{4}{*}{\makecell[l]{Question\\ambiguity}}
            & Entity / column / value mapping & 9 \\
            & Output specification            & 20 \\
            & Aggregation / filter scope      & 15 \\
        \cmidrule(l){2-3}
            & \textbf{Subtotal}               & \textbf{44 (69\%)} \\
        \midrule
        \multirow{4}{*}{\makecell[l]{Model\\defect}}
            & Schema / value grounding        & 12 \\
            & Ignored external knowledge      & 3 \\
            & Join / output / aggregation     & 5 \\
        \cmidrule(l){2-3}
            & \textbf{Subtotal}               & \textbf{20 (31\%)} \\
        \midrule
        \multicolumn{2}{l}{\textbf{Total}}   & \textbf{64} \\
        \bottomrule
    \end{tabular}
\end{table}

\minihead{The residual gap is dominated by question ambiguity, not SQL defects.}
As summarized in Table~\ref{tab:residual}, 44 of the 64 cases (69\%) arise from
question ambiguity: the model selects a defensible reading of an
under-specified question that differs from the gold's. Only 20 cases (31\%) are
genuine model defects. This indicates that \sn-K2.6's SQL-generation competence 
is largely saturated on Arcwise-Plat, and that the residual measured error is 
dominated by the interpretation of ambiguous natural language rather than by the 
model's ability to write correct SQL.

\minihead{Model defects are stable across input noise, while ambiguity is not.}
The genuine model defects are evenly split between the two benchmarks (10 on
Arcwise-Plat-Full and 10 on Arcwise-Plat-SQL), whereas Arcwise-Plat-SQL carries
more ambiguity-driven failures than Arcwise-Plat-Full (24 vs.\ 20). Because
Arcwise-Plat-SQL preserves the original noisy questions and external knowledge
while Arcwise-Plat-Full rewrites them, this is exactly the pattern expected if
the model's intrinsic SQL competence is fixed and the additional Plat-SQL
failures come from unresolved question ambiguity. It also explains why
self-consistency, which aggregates over interpretations, recovers human-level
accuracy on the noisier Arcwise-Plat-SQL (Section~\ref{sec:eval-headline}).

\minihead{When the model fails, it under-constrains.}
Of the 20 model defects, 12 are \emph{schema/value-grounding} errors in which the
model omits an existence or validity filter the schema requires, leaving 3 cases
of ignoring otherwise-usable external knowledge and 5 join/output/aggregation
errors. The model rarely produces logically scrambled SQL. In contrast, when the
model fails, it typically writes a locally reasonable query that fails to encode 
a schema-specific constraint.

\subsection{Question Ambiguity (44 cases)}

\minihead{Entity, column, or value mapping (9 cases).}
The question does not uniquely determine which schema element it refers to. In
Q39, ``Fresno schools'' is graded against \texttt{County = 'Fresno'} while the
model filters \texttt{City = 'Fresno'}. Fresno is both a city and a county. In
Q892, ``the driver with the most points scored'' is computed by the gold over
the per-race \texttt{results.points} column and by the model over the cumulative
\texttt{driverStandings.points} column. The schema exposes two columns named
\texttt{points} with different semantics.

\minihead{Output specification (20 cases).}
The question leaves the exact output under-determined, such as which columns to 
return, in what form, or how a value should be matched or scaled. In some cases, 
the model's choice differs from the gold's. In Q587, ``posts tagged humor'' is 
matched by the gold with exact string equality (\texttt{Tags = '<humor>'}) and 
by the model with substring membership (\texttt{Tags LIKE '\%<humor>\%'}). The 
two differ on posts that carry additional tags. In Q894, ``list the recorded lap 
time'' returns the human-readable \texttt{time} string in the model's query and 
the raw \texttt{milliseconds} integer in the gold. In Q1149, the question asks 
for a \emph{ratio} of male in-patients to outpatients. The gold reports it 
scaled to a percentage while the model returns the unscaled ratio.

\minihead{Aggregation or filter scope (15 cases).}
The question leaves the set being aggregated, or the handling of ties and
multiplicities, open to reading. In Q197, ``single-bonded molecules'' is read by
the gold as molecules with \emph{at least one} single bond and by the model as
molecules whose bonds are \emph{all} single. In Q1080, ``players who prefer the
left foot \ldots\ and have at least one record indicating low work rate'' is
encoded by the gold as both conditions holding in the \emph{same} attribute
record, while the model, following the ``at least one record'' phrasing, matches
the two conditions across different records of the same player.

\subsection{Model Defects (20 cases)}

\minihead{Schema/value grounding (12 cases).}
The model omits a constraint the schema requires. In Q11, ``codes of the
schools'' is graded against a query that filters \texttt{School IS NOT NULL}
because the \texttt{frpm} table contains non-school rows; the model omits the
filter and over-returns---the same omission the case study in
Appendix~\ref{appendix:arcwise-case-study} illustrates.
\begin{lstlisting}
-- gold: restricts frpm to real schools
... WHERE ("Enrollment (K-12)" + "Enrollment (Ages 5-17)") > 500
    AND School IS NOT NULL
-- model: omits the existence filter, over-returns non-school rows
SELECT CDSCode FROM frpm
WHERE ("Enrollment (K-12)" + "Enrollment (Ages 5-17)") > 500;
\end{lstlisting}
In Q1531, the model generates a query that computes average price per item as
\texttt{SUM(Price)/SUM(Amount)}, not recognizing that \texttt{Price} is a
per-unit value, so the schema-correct expression is
\texttt{SUM(Price*Amount)/SUM(Amount)}.

\minihead{Other join/output/aggregation logic (5 cases).}
These include an incorrect inner join that drops valid rows where a left join is
required (Q773) and a query that omits a column the question requests (Q1235).

\minihead{Ignored external knowledge (3 cases).}
In Q407, the evidence maps ``all types of cards'' to the \texttt{subtypes} and
\texttt{supertypes} columns, but the model instead returns the separate
\texttt{type} column, disregarding an explicit and usable hint.

\subsection{The Full--SQL Contrast}

In this section, we present two problems to highlight concretely why Arcwise-Plat-SQL 
is the harder benchmark, by changing character between the two variants of the 
\emph{same} problem. In Q48, the Arcwise-Plat-Full evidence states the ratio's 
denominator explicitly. A model that nonetheless narrows that denominator is 
committing a genuine ignored-evidence defect. In Arcwise-Plat-SQL the same 
problem's evidence omits that formula, so the denominator's scope is genuinely 
ambiguous and the model's reading becomes defensible. In Q1486, Arcwise-Plat-Full 
asks ``how much more do SMEs pay,'' whereas Arcwise-Plat-SQL preserves the 
original ``is it true that more SMEs pay\ldots if so, how many more,'' which 
invites a differently shaped answer. In both, the additional difficulty of 
Arcwise-Plat-SQL is attributable to preserved question ambiguity.

\section{Statistical Methodology}
\label{appendix:stats}

This appendix details how we compute the confidence intervals and significance
tests reported in Section~\ref{sec:eval-headline}. All procedures operate on
the per-instance binary correctness of each method under set-based execution
accuracy, which we release together with our code. Each Arcwise-Plat benchmark
contains $n=498$ instances, and a method's accuracy is the mean of $n$
independent Bernoulli outcomes.

\minihead{Confidence intervals.}
Because the accuracies we report are high ($\hat p \gtrsim 0.9$), the normal
(Wald) approximation has poor coverage near the boundary at this sample size.
We therefore report the Wilson score interval~\cite{wilson1927,brown2001},
which is well calibrated for proportions close to $1$ and moderate $n$. For an
observed accuracy $\hat p = x/n$ at confidence level $1-\alpha$ with
$z = z_{1-\alpha/2}$ (here $\alpha = 0.05$, $z \approx 1.96$), the interval is
\begin{equation}
\frac{\hat p + \dfrac{z^2}{2n} \;\pm\; z\sqrt{\dfrac{\hat p(1-\hat p)}{n} + \dfrac{z^2}{4n^2}}}{1 + \dfrac{z^2}{n}}.
\label{eq:wilson}
\end{equation}
For example, the greedy-decoding accuracy of $93.8\%$ on Arcwise-Plat-Full
yields the $95\%$ interval $[91.3, 95.6]$.

\minihead{Comparison against the human proxy.}
We treat BIRD's reported human execution accuracy, $p_H = 92.96\%$
\cite{bird-leaderboard}, as a fixed reference and use a one-sided exact
binomial test. For each method we test $H_0\!:\, p \geq p_H$ against
$H_1\!:\, p < p_H$ and call the method \emph{significantly below human parity}
when we reject $H_0$ at $\alpha = 0.05$. Every prior system we evaluate is
significantly below $p_H$ ($p < 10^{-4}$); the closest, OpenSearch on
Arcwise-Plat-Full at $88.2\%$, gives $p \approx 1.4\times10^{-5}$. For
\sn-K2.6, the same test fails to reject $H_0$ in every configuration
($p > 0.05$): its accuracy is not significantly below $p_H$, and equivalently
$p_H$ lies within the two-sided $95\%$ Wilson interval of
Eq.~\ref{eq:wilson}. We therefore report \sn-K2.6 as statistically
indistinguishable from the human proxy, and as the first \ts system for which
this holds while all prior systems remain significantly below it.

\section{VeriEQL Configuration}
\label{appendix:verieql}

We use VeriEQL~\cite{he2024verieql}, a bounded SQL equivalence verifier, in two
places: to measure the false-positive rate of result-based rewards during
diagnosis (\S\ref{subsec:diagnosis}), and as the verification-aware component of
the \sn reward during training (\S\ref{subsec:shaping}). In both, VeriEQL is a
\emph{secondary} check layered on top of result-based grading, and it is disabled
at inference time. This appendix specifies its configuration.

\minihead{When VeriEQL is invoked.}
During training, VeriEQL is called only when a 
candidate query's execution result already 
matches the gold result. Its role is to test whether the two queries are \emph{formally}
equivalent, catching cases where the results coincide on the particular database
instance without the queries being semantically equivalent (Failure Mode~1,
\S\ref{subsec:diagnosis}).

\minihead{Solver settings.}
We run VeriEQL with a bounded-verification bound of 2 tuples per table. Each check runs with a 30-second timeout during diagnosis and training.

\minihead{Schema coarsening.}
We extract each table's schema automatically from the SQLite database file and
map every column to one of VeriEQL's two supported types: 
\texttt{INT}, and \texttt{VARCHAR}.
This coarsening, together with the bounded verification, means VeriEQL may return
an inconclusive verdict, which we handle conservatively (\S\ref{subsec:shaping}).

\section{Prompts and Rollout Specification}
\label{sec:app-prompts}

This appendix specifies the exact prompts and rollout mechanics used for \rvbase
and \sn, during both RLVR training and inference. \rvbase uses the base system
prompt of \S\ref{subsec:app-system-prompt}, while \sn extends it with the two
external-knowledge use directives of \S\ref{subsec:app-sn-prompt}. The
\sn's prompt is also what we give the pre-RLVR base model and any single-model 
baseline that does not publish a system prompt of its own 
(Section~\ref{sec:eval-settings}).

\lstdefinestyle{promptstyle}{
    basicstyle=\ttfamily\scriptsize,
    breaklines=true,
    breakatwhitespace=false,
    columns=fullflexible,
    keepspaces=true,
    showstringspaces=false,
    frame=none,
    language=,
    morekeywords={},
}

\subsection{Base System Prompt (\rvbase)}
\label{subsec:app-system-prompt}

We use a universal system prompt for all \ts problems, across all
benchmarks and both SQL dialects. It defines the task, the input format, the
output contract, and the multi-turn protocol.

\begin{tcolorbox}[
        colback=gray!5!white,
        colframe=gray!20!white,
        arc=4pt,
        boxrule=1pt,
        left=3pt, right=3pt, top=3pt, bottom=3pt,
        breakable
    ]
\begin{lstlisting}[style=promptstyle]
Task Overview:
You are a data science expert. Below, you are provided with a database schema and a natural language question. Your task is to understand the schema and generate a valid SQL query to answer the question within limited turns. You should breakdown the problem, draft your reasoning process, and generate the solution.

The task you will receive should have the following format:

Database Engine:
{engine}

Database Schema:
<<db_details>>
This schema describes the database's structure, including tables, columns, primary keys, foreign keys, and any relevant relationships or constraints.

External Knowledge:
<<external_knowledge>>

Question:
<<question>>

Instructions:
- Make sure you only output the information that is asked in the question. If the question asks for a specific column, make sure to only include that column in the SELECT clause, nothing more.
- The generated query should return all of the information asked in the question without any missing or extra information.
- If the external knowledge is not empty, you must use all the information provided in the external knowledge to help you generate the SQL query.
- Before generating the final SQL query, please think through the steps of how to write the query. It should include detailed considerations such as analyzing questions, summarizing relevant findings, brainstorming new ideas, verifying the accuracy of the current steps, refining any errors, thinking of how to call SQL tools, and revisiting previous steps.

Format:
- Conduct thinking every time you get new observation or information.
- You can use the SQL tool to explore data or verify result. You will receive the execution results or error information of the SQL from a user. Based on this information, you can think again and refine.
- The returned dataframe will be truncated in 50 rows if observation is too long.
- Only if you find no further exploration is needed or reach max turns, you directly provide the final SQL query solution inside <solution>...</solution>.
- Do not request a SQL tool execution and provide a solution in the same response.
\end{lstlisting}
\end{tcolorbox}

\subsection{Evidence-Structured Prompt (\sn)}
\label{subsec:app-sn-prompt}

\sn uses the same system prompt with two additional directives in the
\texttt{Format} section, which operationalize the process supervision discussed in 
\S\ref{subsec:shaping}. Every other part of the prompt is identical to 
\S\ref{subsec:app-system-prompt}.

\begin{tcolorbox}[
        colback=gray!5!white,
        colframe=gray!20!white,
        arc=4pt,
        boxrule=1pt,
        left=3pt, right=3pt, top=3pt, bottom=3pt,
        breakable
    ]
\begin{lstlisting}[style=promptstyle]
- In the first or second turn, you must perform a translation from the external knowledge to requirements of the final SQL query. The external knowledge may contain multiple pieces of information separated by semicolons. You must list the requirement for each external knowledge piece the following format: "Requirement of external knowledge {i} ({raw external knowledge text}): ...". This will help you better understand the question and the external knowledge, and guide your thinking in the subsequent turns. If the external knowledge is empty, skip this translation step.

- In the response where you produce the final solution inside <solution>...</solution>, you must first verify the external knowledge one by one and conduct a final check to ensure all the external knowledge is correctly used in the generated SQL query. You must use the following format to verify each piece of external knowledge: "Verification of external knowledge {i} ({raw external knowledge text}): ...". If you find any issues during the verification, you must update your draft before finalizing the query inside <solution>. If the external knowledge is empty, skip this verification step.
\end{lstlisting}
\end{tcolorbox}

\minihead{Relation to process supervision.}
The \emph{requirement} directive forces the model to translate each
external-knowledge entry into an explicit query constraint before it writes SQL.
The \emph{verification} directive forces it to audit the generated query against
each entry, and to revise the draft if the audit fails, in the same response that
commits the answer. Because both blocks are emitted in a fixed, indexed format,
we check the compliance by a string-level test on the reasoning trace, with no
learned reward model in the loop. \S\ref{subsec:app-reward} specifies the
resulting rule set $\mathcal{K}$ and its penalties.

\subsection{Input Format}
\label{subsec:app-input}

Each problem is rendered into the four fields declared by the system prompt.

\begin{enumerate}[leftmargin=*]
\item \textbf{Database Engine.} The target dialect (\texttt{SQLite} or
\texttt{Snowflake}), which fixes the syntax the model must emit.
\item \textbf{Database Schema.} The full schema in data definition language,
including tables, columns, primary keys, and foreign keys. Following prior
work~\cite{li2025omnisql,liu2025skyrlsql}, we annotate each column with the
column description provided by BIRD; when no description is available, we instead
attach three randomly sampled values from that column, which lets the model infer
the value format without querying the database.
\item \textbf{External Knowledge.} The evidence entries associated with the
problem, verbatim. Multiple entries are separated by semicolons, and this
ordering defines the index $i$ used by the requirement and verification blocks.
Problems that carry no external knowledge leave this field empty.
\item \textbf{Question.} The natural language question, verbatim.
\end{enumerate}

\subsection{Rollout Specification}
\label{subsec:app-rollout}

A rollout is a multi-turn interaction between the model and a database
environment. The model is given a tool, \texttt{execute\_sql\_query},
which executes a SQL query against the problem's database and returns the result.

\minihead{Intermediate turns.}
On each intermediate turn the model emits its reasoning followed by a tool call.
The environment executes the query with a 30-second timeout and replies with one
of the following state.
\begin{enumerate}[leftmargin=*]
\item \emph{Success.} The result table, rendered as text. To bound the context,
the table is truncated to its first 50 rows and individual cells are truncated to
200 characters. We explicitly tell the model when truncation occurred.
\item \emph{Execution error.} The database error message, so that the model can
diagnose and repair its query.
\item \emph{Malformed action.} If the tool call cannot be parsed, the environment
returns an instruction to re-emit a tool call.
\end{enumerate}
To aid planning, the environment appends a turn counter to every successful
response: \texttt{``You have \{k\} turns left to complete the task.''} The model
may not request a tool execution and submit a final solution in the same
response.

\minihead{Terminal turn.}
The model terminates a rollout by emitting its final query inside
\texttt{<solution>...</solution>}, which it is instructed to do only when further
exploration is unnecessary or the turn budget is exhausted. A response that
violates the output contract receives a reward of $-1$.

\subsection{Reward Specifications}
\label{subsec:app-reward}

\begin{table}[t]
    \centering
    \footnotesize
    \caption{Prompt and reward configuration of each method. The pre-RLVR base
    model receives the same evidence-structured prompt as
    \sn. \rvbase, our general recipe, uses the
    base prompt and a purely result-based reward. \sn couples the
    evidence-structured prompt with the penalties that enforce it.}
    \label{tab:config-summary}
    \begin{tabular}{l|c|cc}
        \toprule
        & \makecell{System\\prompt}
        & \makecell{Verification-aware\\reward ($\beta$)}
        & \makecell{Process\\penalties ($\lambda_k$)} \\
        \midrule
        Base model  & Evidence & \multicolumn{2}{c}{\textit{not trained}} \\
        \rvbase                & Base     & \myempty & \myempty \\
        \sn                    & Evidence & \myblock & \myblock \\
        \bottomrule
    \end{tabular}
\end{table}

Table~\ref{tab:config-summary} summarizes which prompt and which reward terms each
configuration uses. \rvbase is trained with the base prompt and a purely
result-based reward; the verification-aware stage and the process penalties are
specific to \sn. The reward for a rollout is computed as follows.

\begin{enumerate}[leftmargin=*]
\item \textbf{Format check.} A solution that does not satisfy the output contract
receives $-1$.
\item \textbf{Result-based grading.} We execute the generated query and the gold
query against the database and compare their results using the grading method
annotated for that instance---set-based, subset-based, or list-based
(\S\ref{subsec:errors}). A mismatch yields reward $0$. This is the only reward
term used by \rvbase.
\item \textbf{Verification-aware grading (\sn only).} If result-based grading
accepts the query, we invoke VeriEQL~\cite{he2024verieql} with a 30-second 
timeout to test semantic equivalence to the gold query. If VeriEQL refutes 
equivalence, the reward is downweighted to $1-\beta = 0.8$; if VeriEQL confirms 
equivalence, times out, or does not support the query, the reward is $1$. This 
stage is applied to SQLite problems, the dialect VeriEQL supports.
\item \textbf{Process penalties (\sn only).} We deduct $\lambda_k = 0.1$ for each
violated rule $k \in \mathcal{K}$, as specified below.
\end{enumerate}

\minihead{Evidence-supervision rules.}
The rule set $\mathcal{K}$ of Equation~\ref{eq:revisql-reward} contains two rules,
each carrying $\lambda_k = 0.1$. Both are evaluated by string-level checks on the
reasoning trace.
\begin{enumerate}[leftmargin=*]
\item \textbf{Requirement rule.} The requirement block must appear by the second
turn. We evaluate this rule once per rollout, at turn two, and deduct $0.1$ if the
block is absent. Because it is checked before the model commits an answer, the
penalty applies whether or not the rollout ultimately produces a solution.
\item \textbf{Verification rule.} The verification block must appear in the same
response as the final \texttt{<solution>}. We evaluate this rule when the model
commits a solution and deduct $0.1$ if the block is absent.
\end{enumerate}
We relax both rules in a special case when the model answers on the first
turn, which happens on easy problems that require no exploration. In this case, 
demanding a separate translate-then-verify pass is unnecessary, so we deduct 
$0.1$ only if \emph{both} blocks are missing, and nothing if either is present. 
A rollout therefore incurs at most $0.2$ in process penalties, and at most $0.1$ 
when it terminates on its first turn. Problems that carry no external knowledge 
require neither block.

\end{document}